\documentclass[aps,superscriptaddress,showpacs,prx,nofootinbib,twocolumn]{revtex4-1}
\usepackage{bm} 
\usepackage{graphicx} 
\usepackage{amsmath}
\usepackage{amsthm}
\usepackage{amssymb} 
\usepackage{epstopdf} 
\usepackage{amsfonts}
\usepackage{epsfig}
\usepackage{color} 
\usepackage{subfigure}
\usepackage[english]{babel}

\newcommand{\bra}[1]{\langle #1 |}
\newcommand{\ket}[1]{| #1 \rangle}

\newcommand {\be}{\begin{equation}}
\newcommand {\ee}{\end{equation}}

\newcommand\D{{\text{c}}}

\newcommand{\ba}{\begin{eqnarray}}
\newcommand{\ea}{\end{eqnarray}}
\newcommand\tr{{\mbox{Tr\,}}}
\newcommand{\ignore}[1]{}

\usepackage[margin=1in,nofoot]{geometry}

\newcommand{\Tr}{{\mathrm{Tr}}}
\newcommand{\e}{{{e}}}
\newcommand{\x}{{E}}
\newcommand{\rmd}{{\text d}}

\newcommand{\beq}{\begin{equation}}
\newcommand{\eeq}{\end{equation}}
\newcommand{\beqnn}{\begin{equation*}}
\newcommand{\eeqnn}{\end{equation*}}
\newcommand{\bea}{\begin{eqnarray}}
\newcommand{\eea}{\end{eqnarray}}
\newcommand{\beann}{\begin{eqnarray*}}
\newcommand{\eeann}{\end{eqnarray*}}
\newcommand{\bes} {\begin{subequations}}
\newcommand{\ees} {\end{subequations}}

\begin{document}

\title{Off-Diagonal Expansion Quantum Monte Carlo}
\author{Tameem Albash}
\affiliation{Information Sciences Institute, University of Southern California, Marina del Rey, California 90292, USA}
\affiliation{Department of Physics and Astronomy and Center for Quantum Information Science \& Technology, University of Southern California, Los Angeles, California 90089, USA}
\author{Gene Wagenbreth}
\affiliation{Cray, Seattle, Washington 98164, USA}
\author{Itay Hen}
\email{itayhen@isi.edu}
\affiliation{Information Sciences Institute, University of Southern California, Marina del Rey, California 90292, USA}
\affiliation{Department of Physics and Astronomy and Center for Quantum Information Science \& Technology, University of Southern California, Los Angeles, California 90089, USA}

\date{\today}

\begin{abstract}
We propose a Monte Carlo algorithm designed to simulate quantum as well as classical systems at equilibrium, bridging the algorithmic gap between quantum and classical thermal simulation algorithms.  The method is based on a novel decomposition of the quantum partition function that can be viewed as a series expansion about its classical part.  We argue that the algorithm is optimally suited to tackle quantum many-body systems that exhibit a range of behaviors from `fully-quantum' to `fully-classical', in contrast to many existing methods.  We demonstrate the advantages of the technique by comparing it against existing schemes.  We also illustrate how our method allows for the unification of quantum and classical thermal parallel tempering techniques into a single algorithm and discuss its practical significance. 
\end{abstract}

\maketitle

\section{Introduction}

Quantum Monte Carlo (QMC) algorithms are known to be notoriously inefficient in `almost classical' parameter regimes where updates resulting from thermal fluctuations are expected to be far more dominant than those resulting from quantum fluctuations.  This is particularly true in models that can be parametrically tuned from quantum to classical regimes, such as the transverse-field Ising model~\cite{tfim1,tfim2,Chakrabarti:1980}, the XXZ model~\cite{Lieb:61,pasquale:2008} or the Bose-Hubbard model~\cite{Kuhner:1998,jaksch,Gersch:1963}\footnote{The Bose-Hubbard models exhibits a quantum phase transition from a highly delocalized superfluid at one extreme of its parameter space to a classical localized Mott insulator at the other extreme.}.  Since QMC methods evolve via configuration updates that are based on quantum fluctuations, the acceptance rates of quantum updates, e.g., single spin (or cluster) flips in the Ising system, decrease dramatically in classical regimes, often causing QMC algorithms to dramatically slow down or `freeze'~(see, e.g., Ref~\cite{farhi:12}).  

Efficient classical thermal updates are typically hard to implement within the framework of QMC algorithms because these algorithms do not normally converge to classical Monte Carlo algorithms in the limit where the model becomes classical.  For this reason, there are almost no algorithms that efficiently simulate systems that exhibit the full range of behavior from being `fully-quantum' to `fully-classical'. 
For the successful simulation of systems exhibiting the above characteristics, it is therefore important to devise Monte Carlo schemes that can function both as quantum as well as classical algorithms when necessary. Efficient methods of this type will have wide-ranging applicability in diverse areas ranging from statistical physics through quantum chemistry to quantum computing, to mention a few areas.

Here, we propose an algorithm that has the algorithmic flexibility to simulate interacting many-body systems ranging from the fully-quantum to the fully-classical.  We present a Monte Carlo scheme that is based on a novel decomposition of the canonical quantum partition function into a sum of `generalized' Boltzmann weights and that converges to the usual decomposition of the classical partition function in the limit where the Hamiltonian of the system becomes classical. Based on this unique decomposition, our algorithm aims to improve the convergence rates of simulated systems for which existing techniques are often inefficient. 

Within our approach the quantum imaginary-time dimension of the algorithm is `elastic', i.e., it can stretch or shrink dynamically depending on the strength of the quantum part of the system --- the off-diagonal portion of the Hamiltonian.  In addition, the proposed method does not introduce Trotter-type errors as in path-integral QMC (PIQMC), a source of errors that normally occurs from an insufficient discretization of the imaginary-time dimension (over-discretization tends to sharply reduce the acceptance rates of the QMC updates). Moreover, in the classical limit where off-diagonal terms vanish, our algorithm naturally reduces to a classical thermal simulation.  As we illustrate, the above properties allow our method to naturally overcome certain inefficiencies typically encountered by other QMC algorithms. 

The structure of the paper is as follows.  In Sec.~\ref{eqt:Zexpansion}, we describe the decomposition of the canonical quantum partition function into what we refer to as generalized Boltzmann weights (GBWs).  We then proceed in Sec.~\ref{sec:Algorithm} to present the basic steps, updates and measurements of our off-diagonal expansion (ODE) quantum Monte Carlo algorithm that builds on the above decomposition. To illustrate the practicality of our algorithm, we examine in Sec.~\ref{sec:Results} simulations of the transverse field Ising model, especially inside the spin-glass phase, where the model is known to be hard to simulate. We also discuss in this context the unification of quantum and classical parallel tempering. We present some conclusions in Sec.~\ref{sec:conclusions}. 

\section{Generalized Boltzmann Weights} \label{eqt:Zexpansion}

\subsection{Decomposition of the partition function} 
%
The main insight at the heart of our approach is a novel decomposition of the canonical quantum partition function which, as we argue, allows for the development of a QMC algorithm that has certain advantages over existing methods.  Our work builds in part on the Stochastic Series Expansion (SSE) algorithm, a well-known and successful QMC algorithm pioneered by Sandvik~\cite{sandvik:05,sandvik:10}, which---unlike the traditional `slicing' of the partition function into Trotter segments---involves a Taylor series expansion in the inverse-temperature $\beta=1/T$ (in our units $k_B=1$) of the partition function as was originally suggested by Handscomb~\cite{handscomb1,handscomb2}.  

The canonical quantum partition function of a system described by a Hamiltonian $H$ is given by $Z=\tr \left[ \e^{-\beta H}\right]$.  Our decomposition begins by first writing the Hamiltonian in the form 
\beq
H = H_\D - \sum_j \Gamma_j  V_j \,.
\eeq
Here, $H_\D$ is a `classical' Hamiltonian, i.e., a diagonal operator in some known basis, which we refer to as the computational basis, and whose basis states will be denoted by $\{ |z\rangle \}$. 
The $\{ \Gamma_j \}$ are generally complex-valued parameters, and $\{V_j \}$ are off-diagonal operators satisfying $[V_j,H_\D] \neq 0$ that give the system its  `quantum dimension'.  In an analogous way to standard SSE, in order for the decomposition of the partition function to be feasible, we require the off-diagonal operators to be chosen such that they obey 
${V}_j | z \rangle = | z' \rangle$
for every basis state $|z\rangle$, where $ | z' \rangle \neq |z\rangle$ is also a basis state.
For simplicity we henceforth assume that all the $\Gamma_j$ parameters are identical, namely that  $\Gamma_j= \Gamma , \forall j$, however as will become evident shortly this restriction is by no means necessary. 

We now present the main steps for the decomposition of the quantum partition function. 
We first replace the trace operation $\Tr[\cdot ]$ with the explicit sum $\sum_z \langle z | \cdot | z \rangle$ and then expand the exponent in the partition function in a Taylor series:
\bes
\begin{align}
Z =&\sum_{z} \sum_{n=0}^{\infty}\frac{\beta^n}{n!} \langle z | (-H)^n | z \rangle \\
=& \sum_{z} \sum_{n=0}^{\infty}\frac{\beta^n}{n!} \langle z | (-H_\D + \Gamma \sum_j V_j)^n | z \rangle \\
=& \sum_{z} \sum_{n=0}^{\infty}  \sum_{\{ {S}_{n}\}} \frac{\beta^n}{n!} \langle z | {S}_{n} | z \rangle \,,
\end{align}
\ees
where in the last step we have also expanded $(-H)^n$, and $\{{S}_{n}\}$ denotes the set of all sequences  of length $n$ composed of products of basic operators $H_\D$ and $V_j$.

\begin{figure}[htp]
\includegraphics[trim={4cm 10.5cm 17cm 3cm},clip,width=0.9\columnwidth]{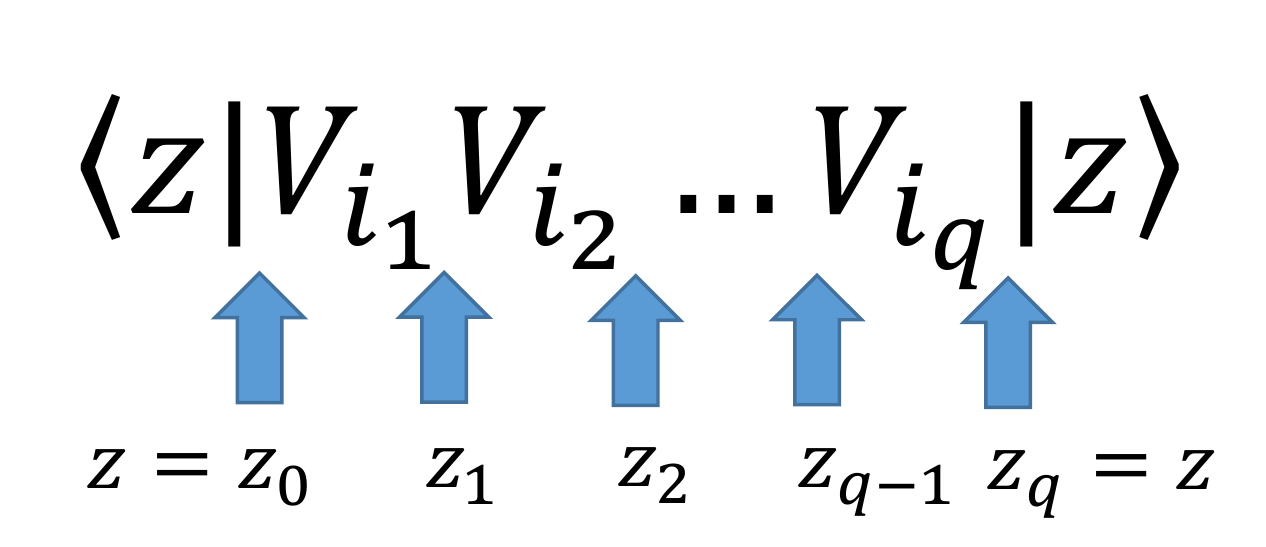}
\caption{\label{fig:fig3} {\bf A diagrammatic representation of the term $\langle z| S_q|z\rangle$.} The sequence of operators \hbox{${S}_{q}={V}_{i_1} \cdot {V}_{i_2} \cdots {V}_{i_q}$} is sandwiched between classical bra $\langle z|$ and ket $|z\rangle$, inducing a sequence of classical states ($|z_0\rangle, \ldots, |z_q\rangle$). The multiset of classical energies of the states $|z_i\rangle$, namely,
$E_i=E(z_i)=\langle z_i | H_\D|z_i\rangle$ generate the generalized Boltzmann weight.} 
\label{fig:sequence}
\end{figure}

We proceed by removing all the diagonal Hamiltonian terms from the sequence $\langle z | {S}_n | z \rangle$. We do so by evaluating their action on the relevant basis states, leaving only the off-diagonal operators unevaluated inside the sequence. 
At this point, the partition function can be written as: 
\bea\label{eq:snsq}
Z  &=&
\sum_{z} \sum_{q=0}^{\infty} \sum_{\{ {S}_{q}\}}  \Gamma^q \langle z | {S}_q | z \rangle  \left( \sum_{n=q}^{\infty} \frac{\beta^n(-1)^{n-q}}{n!} \right. \nonumber \\ 
&\times&\left.  \sum_{\sum k_i=n-q} E^{k_0}(z_0) \cdot \ldots \cdot E^{k_{q}}(z_{q}) \right)\,,
\eea
where $E(z_i)=\langle z_i |H_\D | z_i \rangle$ and ${\{{S}_q\}}$ denotes the set of all sequences of length $q$ of `bare' \emph{off-diagonal} operators $V_j$. The term in parenthesis sums over the diagonal contribution of all $\langle z | {S}_n | z \rangle$ terms that correspond to a single $\langle z | {S}_q | z \rangle$ term. The various $\{|z_i\rangle\}$ states are the states obtained from the action of the ordered $V_j$ operators in the sequence ${S}_q$ on $|z_0\rangle$, then on $|z_1\rangle$, and so forth\footnote{For example, for ${S}_q={V}_{i_q} \ldots {V}_{i_2}{V}_{i_1}$, 
we obtain $|z_0\rangle=|z\rangle, {V}_{i_1}|z_0\rangle=|z_1\rangle, {V}_{i_2}|z_1\rangle=|z_2\rangle$, etc.}.
Figure~\ref{fig:sequence} gives a schematic representation of $\langle z | {S}_q | z \rangle$. 
After a change of variables, $n \to n+q$, we arrive at:
\bea
Z  &=& \sum_{z} \sum_{q=0}^{\infty} \sum_{\{ {S}_{q}\}} \langle z | {S}_q | z \rangle \left( (\beta \Gamma)^q \sum_{n=0}^{\infty} \frac{(-\beta)^n}{(n+q)!} \right. \nonumber  \\
&\times&
\left. \sum_{\sum k_i=n}  E^{k_0}(z_0) \cdots E^{k_{q}}(z_{q}) \right)\,.
\eea
%
Abbreviating $\x_i \equiv E(z_i)$ (note that the various $\{\x_i\}$ are functions of the $|z_i\rangle$ states created by the operator sequence ${S}_q$), the partition function is now given by:
\bea
Z  &=& \sum_{q=0}^{\infty}(-\Gamma)^q \sum_{z, \{ {S}_{q}\}}  \langle z | {S}_q | z \rangle 
\\\nonumber&\times&
 \left( 
\sum_{\{ k_i\}=(0,\ldots,0)}^{(\infty,\ldots,\infty)} \frac{(-\beta)^q}{(q+\sum k_i)!} \prod _{j=0}^{q} (-\beta \x_j)^{k_j} 
\right)
 \,.
 \eea
A feature of the above infinite sum is that the term in parentheses can be further simplified to give the \emph{exponent of divided differences} of the $\x_i$'s (we give a short description of divided differences and an accompanying proof of the above assertion in Appendix~\ref{sec:DividedDifference}), namely it can be succinctly rewritten as: 
\bea
\sum_{\{ k_i\}} \frac{(-\beta)^q}{(q+\sum k_i)!} \prod _{j=0}^{q} (-\beta \x_j)^{k_j} 
=e^{-\beta[\x_0,\ldots,\x_q]} \,, 
\eea
where $[\x_0,\ldots,\x_q]$ is a \emph{multiset} of energies and where a function $F[\cdot]$ of a multiset of input values is defined by
\beq 
F[\x_0,\ldots,\x_q] \equiv \sum_{j=0}^{q} \frac{F(\x_j)}{\prod_{k \neq j}(\x_j-\x_k)}
\eeq
and is called the \textit{divided differences}~\cite{dd:67,deboor:05} of the function $F[\cdot]$ with respect to the list of real-valued input variables $[\x_0,\ldots,\x_q]$. In our case, $F[\cdot]$ is the function
\beq
F[\x_0,\ldots,\x_q] = e^{-\beta[\x_0,\ldots,\x_q]} \,.
\eeq
A feature of divided differences is that they are invariant under rearrangement of the input values, so the input sequence forms a multiset, i.e., a generalization of the mathematical set which allows repetitions but where order does not play a role.  The partition function in its close-to-final form is thus given by:
\beq
Z  = \sum_{z} \sum_{q=0}^{\infty} \sum_{\{ {S}_{q}\}} \langle z | {S}_q | z \rangle (-\Gamma)^q e^{-\beta[\x_0,\ldots,\x_q]} \,.
\label{eq:SSE}
\eeq
We note that a single divided-difference term is a sum of an infinite number of terms in the usual breakdown of SSE.  This can be immediately seen in Eq.~\eqref{eq:snsq}, which relates the standard SSE weight, involving sequences of diagonal as well as off-diagonal bonds (denoted by ${S}_n$), to the weights of the current algorithm that only involve off-diagonal bonds\footnote{We note here however that the computational cost of calculating the weights of configurations here is higher than in standard SSE, due to the need to evaluate the divided differences of the exponential function over the energies of the configuration. As we show below, the cost of each such evaluation can be shown to be proportional in the worst case to the square of the number of terms in the sequence ${S}_q$ (by direct divided-differences calculation, see Appendix~\ref{sec:DividedDifference} and, e.g., Ref.~\cite{dd:67}), which scales linearly with the inverse temperature $\beta$ and with the number of particles in the system $N$ (this is discussed in detail later on).}.  

Furthermore, the mean value theorem for divided differences~\cite{dd:67,deboor:05} together with the monotonicity of the exponential function allows us to write   
\bea
\e^{-\beta[\x_0,\ldots,\x_q]} & =&\frac{\rmd^{q} \left(\e^{-\beta \x}\right)}{\rmd \x^q}\Bigg|_{\x=\x_{(0,\ldots,q)}}  \\\nonumber
& =&\frac{(-\beta)^q \e^{-\beta \x_{(0,\ldots,q)}}}{q!}  \,,
\eea
for a single real-valued energy \hbox{$\x_{(0,\ldots,q)} \in\left( \min[\x_0,\ldots,\x_q],\max[\x_0,\ldots,\x_q]\right)$} calculated from the multiset $[\x_0,\ldots,\x_q]$ and $\beta$. Specifically, $\x_{(0,\ldots,q)}$ lies within the spectrum of the classical Hamiltonian.
This allows us to write the partition function in terms of classical `effective energies' as 
 \beq
Z  =\sum_{\{z\}} \sum_q  \sum_{\{S_q\}}  \frac{(\beta \Gamma)^q}{q!} \bra{z} S_q \ket{z}  \e^{-\beta \x_{(0,\ldots,q)}} \, .
\eeq
To calculate $\x_{(0,\ldots,q)}$, one may use the divided differences recursion relation (see Appendix~\ref{sec:DividedDifference})
\bea\label{eq:ddr2}
&&F[\x_i,\ldots,\x_{i+j}] \\\nonumber
&=& \frac{F[\x_{i+1},\ldots , \x_{i+j}] - F[\x_i,\ldots , \x_{i+j-1}]}{\x_{i+j}-\x_i} \,,
\eea 
which in terms of effective classical energies becomes
\bea
\frac{(-\beta)^q}{q!}\e^{-\beta \x_{(0,\ldots,q)}} &=& \nonumber \\
&& \hspace{-2.5cm} \frac{(-\beta)^{q-1}}{(q-1)!}
\frac{
\left(\e^{-\beta \x_{(0,\ldots,q-1)}} - \e^{-\beta \x_{(1,\ldots,q)}}\right)
}{\x_0-\x_q} \,.
\eea
Isolating $\x_{(0,\ldots,q)}$, we arrive at
\beq \label{eqt:EffectiveE}
\x_{(0,\ldots,q)}  = \bar{\x} -\frac1{\beta}\log \frac{2 q \sinh \beta \Delta \x}{\beta(\x_q-\x_0)} \ ,
\eeq
where 
\bea
2\bar{\x} &=& \x_{(1,\ldots,q)}+\x_{(0,\ldots,q-1)} \quad \text{and} \nonumber \\
2 \Delta \x &=&\x_{(1,\ldots,q)} - \x_{(0,\ldots,q-1)}\,. \nonumber
\eea
In the limiting case where all energies in the sequence are equal, the above relation neatly becomes $\x_{(0,\ldots,q)}=\x_{(0)}=\x_0$. 
The initial condition for the above recursion is simply $\e^{-\beta \x_{(i)}} =\e^{-\beta \x_i}$. An illustration of how the recursion relation is used to calculate $\x_{(0,\ldots,q)}$ is depicted in Fig.~\ref{fig:pyr1} and is discussed in more detail in Appendix~\ref{sec:GBWdetails}.

\begin{figure}[th]
\includegraphics[width=0.49\textwidth]{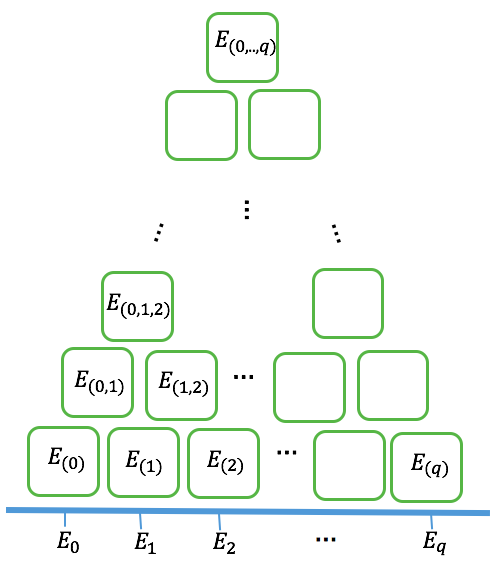}
\caption{ {\bf Calculating the effective classical energy of the generalized Boltzmann weight using a `pyramid' structure.}   The evaluation of the divided differences of the exponential function of $q+1$ input energies consists of calculating each level of the pyramid starting at its base.  The values at the base of the pyramid $E_{(j)}$ are simply the energy inputs $E_j$ (shown as the blue line at the bottom of the pyramid), with all identical energies placed together as a group (the exact ordering of the energies is not important).  To calculate the elements at the next level of the pyramid, we use the relation in Eq.~\eqref{eqt:EffectiveE}.  This procedure is continued until the final level of the pyramid is evaluated, which corresponds to the effective classical energy in the GBW, namely, $E_{(0,\ldots,q)}$.} 
\label{fig:pyr1}
\end{figure}

Finally, since by construction the term $\langle z | {S}_q | z \rangle$ evaluates to either $0$ or to $1$ (the operation $S_q|z\rangle$ returns a basis state $|z'\rangle$ and therefore $\langle z | S_q |z\rangle=\langle z | z'\rangle=\delta_{z,z'}$), the partition function can be rewritten in its final form as a sum over only non-vanishing terms:
 \beq
Z  =\sum_{\{{S}_{q} : \langle z | {S}_q | z \rangle=1\}}  \frac{(\beta \Gamma)^q}{q!} \e^{-\beta \x_{(0,\ldots,q)}}
 \,.
\label{eq:SSE2}
\eeq
We interpret the terms in the sum in Eq.~\eqref{eq:SSE2} as weights, i.e., $Z  = \sum_{\{\mathcal{C}\}} W_{\mathcal{C}}$, where the set of configurations $\{\mathcal{C}\}$ is all the distinct pairs $\{ |z\rangle, {S}_q \}$.  Because of the form of $W_{\mathcal{C}}$, 
\beq \label{eq:gbw}
W_{\mathcal{C}}= \frac{(\beta \Gamma)^q}{q!}   \e^{-\beta E_{(0,\ldots,q)}}\, , 
\eeq
we refer to it as  a `generalized Boltzmann weight' (or, GBW).  We shall refer to $E_{(0,\ldots,q)}$ as the `effective classical energy' of the configuration ${\mathcal{C}}$ and denote it at times for brevity simply by $E_{\mathcal{C}}$.

In order to interpret the $W_{\mathcal{C}}$ terms as actual weights, they must be non-negative for any simulated system that is not plagued by the sign problem~\cite{newman}. It is therefore interesting to note that the above weights are automatically positive if $\Gamma$ is positive, i.e., if the off-diagonal elements are non-positive, which is the case for the so-called stoquastic Hamiltonians~\cite{Bravyi:QIC08,Bravyi:2014bf}. As is also evident from the above expression, even values of $q$ also yield positive weights regardless of the sign of $\Gamma$. This corresponds to a scenario where off-diagonal operators must be injected along the imaginary time dimension in pairs in order to ensure nonzero weights. One such example is the transverse-field Ising Hamiltonian 
\beq\label{eq:Hising}
H =\sum_{\langle i,j\rangle} J_{ij} \sigma^z_i \sigma^z_j +  \sum_j h_j \sigma^z_j - \Gamma \sum_j \sigma^x_j  \, ,
\eeq
where \hbox{$H_\D= \sum_{\langle i,j\rangle} J_{ij} \sigma^z_i \sigma^z_j +  \sum_j h_j \sigma^z_j$} and the off-diagonal operators are the spin-flip terms $V_j = \sigma_j^x$.  In order for the $\langle z| {S}_q|z\rangle$ terms to evaluate to one rather than to zero, off-diagonal operators must always be produced and annihilated in pairs, implying that the total sign of the weight, Eq.~(\ref{eq:gbw}), is positive.
We have thus established a decomposition of the canonical quantum partition function into a sum of positive-valued weights. 
 
\subsection{Properties of the GBWs}
%
One property of the above decomposition of the canonical quantum partition function is that it may be viewed as unifying the classical and quantum partition functions. Specifically, it contains as a sub sum the classical partition function decomposition of its diagonal part $H_\D$.  Writing the quantum partition function as a series in the `quantum strength' parameter $\Gamma$, one obtains the classical partition function as the zeroth term, namely, 
\bea
Z&=&\sum_{\{ {S}_{0} : \langle z | z \rangle =1\}} \e^{-\beta E(z)} \nonumber\\
&+& \frac{(\beta \Gamma)^2}{2} \sum_{\{ {S}_{2} : \langle z | {S}_2 | z \rangle =1\}} \e^{-\beta E_{(0,1,2)}}  + \ldots 
\eea
%
Furthermore, in classical regimes where $\Gamma$ is zero or very small, the dominant configurations, i.e., those with highest weights, 
have no off-diagonal terms, and only the $q=0$ terms survives. In this case the typical weights are 
\beq
\frac{(\beta \Gamma)^q}{q!} \e^{-\beta E_{(0,\ldots,q)}} \Big|_{q=0}= \e^{-\beta\x_{(0)}} = \e^{-\beta E(z)} \,,
\eeq
where $E(z)$ is the classical energy of the spin configuration $z$.  Our decomposition thus automatically reduces to the usual sum over Boltzmann weight of classical Hamiltonians\footnote{This is to be contrasted with other decompositions of the partition function where the classical limit is either unnatural or ill-defined.}.


The GBW, Eq.~(\ref{eq:gbw}), also has several attractive properties that make it useful for Monte Carlo simulations.  First, as was already mentioned, it is strictly positive for stoquastic systems. This feature automatically resolves the `diagonal sign problem' that sometimes appears in other schemes~\cite{dorneich:01}, where constants must be added to the diagonal bonds to rectify the problem. Moreover, since the addition of such constants considerably affects the convergence rate of the algorithm~\cite{sandvik:99,dorneich:01,sandvik:03,sandvik:05,sandvik:10}, these constants usually have to be optimized for faster convergence. A QMC algorithm based on the GBW decomposition is in this respect parameter-free, a property that is expected to facilitate computations.

Second, for any arbitrary energy shift $\Delta E$ of the diagonal part of the Hamiltonian, the following holds:
\beq
\e^{-\beta[\x_0+\Delta E,\ldots,\x_q+\Delta E]}=\e^{-\beta \Delta E} \e^{-\beta[\x_0,\ldots,\x_q]}\,.
\eeq
This identity reflects the fact that the addition of constants to the simulated Hamiltonian
has a trivial effect on the various weights. Specifically, ratios of weights, which in turn determine the acceptance rates of the QMC updates, are invariant under the above addition of a constant, as they should be. 

On a more academic note, it is interesting to observe that the proposed algorithm also has close relations to continuous-time QMC methods   (e.g., Ref.~\cite{prokofiev:98}), via the 
Hermite-Genocchi formula~\cite{deboor:05}:
\bea
\e^{-\beta[\x_0,\ldots,\x_q]} = \int_{\Omega} \rmd t_0 \cdots \rmd t_{q} 
\e^{-\beta\left( \x_0 t_0 +\x_1 t_1 + \ldots + \x_q t_q\right) }  \,, \nonumber\\
\eea
where $t_i \geq 0$ and the area of integration $\Omega$ is bounded by $t_0 + t_1 + \ldots + t_{q}$ from above.

\subsection{A simple analytical example}

As a first illustration, let us consider as a simple example the transverse-field Ising Hamiltonian where the classical Ising part vanishes, namely, where $H_{\D}=0$. In this case the model becomes the trivial system $H = -\Gamma \sum_i \sigma^x_i$.
The partition function in this special case is decomposed as:
\beq
Z=\sum_{\{S_q: \langle z | {S}_q | z \rangle =1\}} \frac{(\beta \Gamma)^q}{q!}  \ ,
\eeq
where the classical energies are $\x_0=\cdots=\x_q=0$, corresponding to $\x_{(0,\ldots,q)}=0$. In this case, we have $\langle z | S_q |z\rangle=1$ if and only if all $\sigma^x_i$ off-diagonal operators in $S_q$ appear an even number of times.  Denoting by $N_p(q)$ the number of nonzero weights for each value of (even) $q$ and every $|z\rangle$, the partition function can be simplified to
\beq
Z=2^N \sum_{q \mathrm{\ even}} \frac{(\beta \Gamma)^q}{q!} N_p(q) \,.
\eeq
A simple calculation (see Appendix~\ref{sec:npq}) reveals
\beq\label{eq:npq}
N_p(q)=\frac1{2^N}\sum_{k=0} {N \choose k} (N-2k)^q\,,
\eeq 
which yields
\bea
Z&=& \sum_q \frac{(\beta \Gamma)^q}{q!} \sum_{k=0} {N \choose k}  (N-2k)^q \\\nonumber
&=&  \sum_{k=0} {N \choose k}\sum_{q\geq0, \text{even}} \left[ \beta \Gamma (N-2k) \right]^q
 \,.
\eea
Carrying out the sum over $q$, we end up with:
\beq
Z=\sum_{k=0} {N \choose k} \cosh\left[\beta \Gamma (N-2k)\right]
=\left(2 \cosh \beta \Gamma\right)^N \, ,
\eeq
which is the correct result for the partition function for the non-interacting system $H = -\Gamma \sum_i \sigma^x_i$.

\section{Off-Diagonal Expansion QMC algorithm} \label{sec:Algorithm}



We now describe the basic ingredients of our Off-Diagonal Expansion (ODE) algorithm that is based on the above partition function decomposition. For concreteness we discuss the algorithm as it applies to the transverse-field Ising model, Eq.~(\ref{eq:Hising}), however we note that generalization to other systems should be straightforward. 
We first establish the computational complexity associated with implementing this new algorithm, discussing in detail generic updates as well as measurements.  We then present some results that allow us to fully characterize and to some extent quantify the advantages of the algorithm over generic QMC methods, specifically path-integral QMC.
\subsection{General description of the algorithm}

An ODE configuration is a pair $\mathcal{C}=\{|z\rangle, S_q\}$ where $|z\rangle$ corresponds to a classical bit configuration and 
\hbox{$S_q=V_{i_1} V_{i_2} \cdots V_{i_q}$} is a sequence of (possibly repeated) off-diagonal operators. 
As was discussed above, each configuration $\mathcal{C}$ induces a list of states $Z=\{|z_0\rangle =|z\rangle ,|z_1\rangle ,\ldots,|z_q\rangle =|z\rangle \}$ (see Fig.~\ref{fig:sequence}), which in turn also generates a corresponding {multiset} of diagonal 
energies $M_{\mathcal{C}}=\{E_0,E_1,\ldots,E_q\}$ of not-necessarily-distinct values (recall that $E_i = \langle z_i | H_c | z_i\rangle$). For systems with discretized energy values, the multiset can be stored efficiently in a `multiplicity table' 
$\{ m_0,m_1,\ldots, m_j,\ldots\}$, where $m_j$ is the multiplicity of the energy $E_j$ in the multiset.
Given $M_{\mathcal{C}}$, the evaluation of the effective classical energy $E_{\mathcal{C}}$ and the GBW $W_{\mathcal{C}}$ follow from the definition of the GBW, Eq.~(\ref{eq:gbw}).
The actual evaluation of the effective classical energy is schematically given in Fig.~\ref{fig:pyr1} and is discussed in more technical detail in Appendix~\ref{sec:GBWdetails}. 

The initial configuration of the ODE algorithm is a random classical configuration $|z\rangle$ and the empty sequence $S_{q=0}=1$.
The weight of this initial configuration is 
\beq
W_{\mathcal{C}_{\text{init}}}=\e^{-\beta E(z)} \,,
\eeq
i.e., the classical Boltzmann weight of the initial random state $|z\rangle$. Here the effective classical energy $E_{\mathcal{C}_{\text{init}}}$ is the classical energy of $|z\rangle$. 

\subsection{Updates}



\begin{figure}
\includegraphics[trim={.2cm 1.2cm 4.4cm .5cm},clip,width=0.99\columnwidth]{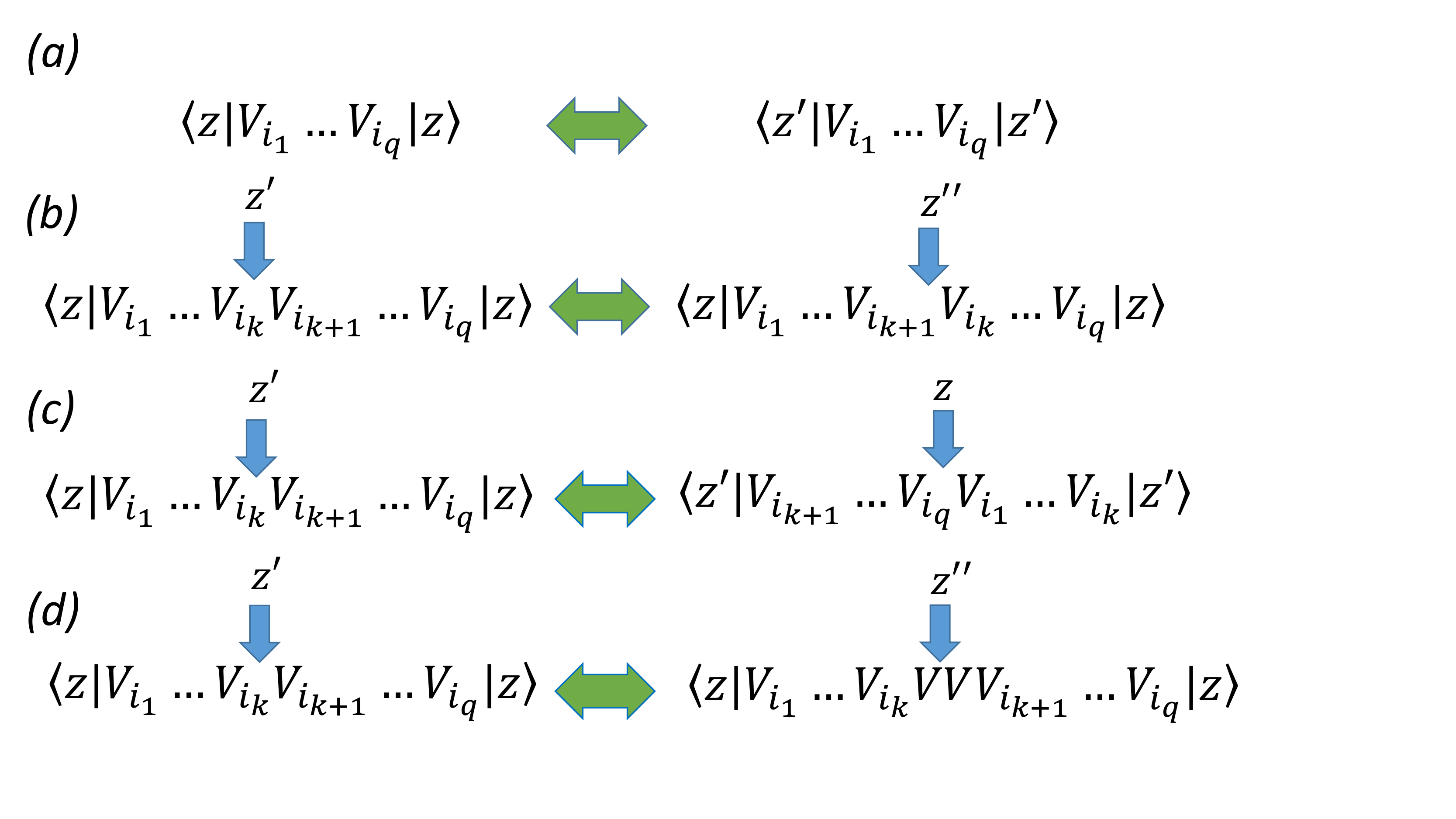}
\caption{{\bf Basic update moves of the ODE algorithm.} (a) Classical moves (e.g., a single bit flip), whereby only the initial state $z$ is changed to $z'$ leaving $S_q$ unchanged. 
(b) Local swap, whereby two adjacent operators $V_{i_k} V_{i_{k+1}}$ are interchanged changing the state between them from $z'$ to $z''$. (c) Block swap, whereby two partitions of the sequence are interchanged.  This changes the initial state from $z$ to $z'$ as well as the ordering of the sequence. (d) Pair creation/annihilation, whereby a new pair of operators is inserted or deleted. 
}
\label{fig:updates}
\end{figure}

We next describe the basic update moves for the algorithm.  We consider here only generic local updates but note that updates of the global type can be specifically tailored to the system in question.  An update is considered local if it changes the multiset $M_\mathcal{C}$ by a finite (i.e., by a system-size independent) number of terms, e.g.,  $M_{\mathcal{C}} \to M_{\mathcal{C}} +\{ E(z_i)\} -\{ E(z_j)\}$. The basic updates are summarized in Fig.~\ref{fig:updates} and are discussed in detail below. 

\subsubsection{Classical moves}
Classical moves are any moves that involve a manipulation of the classical state $|z\rangle$ while leaving $S_q$ unchanged [see Fig.~\ref{fig:updates}(a)]. 
In a single bit-flip classical move, a spin from the classical bit-string state $\ket{z}$ of $\mathcal{C}$ is picked randomly and is flipped, generating a state $\ket{z'}$ and hence a new configuration $\mathcal{C}'$.  Performing this change requires recalculating the energies associated with the sequence $S_q$ leading to a new multiset $M_{\mathcal{C}'}$ and can become computationally intensive if $q$ is large. Classical moves should  therefore be attempted with low probabilities if $q$ large. Simply enough, the acceptance probability for a classical move is
\beq\label{eq:Pmet}
p = \min \left( 1,\frac{W_{\mathcal{C}'}}{W_{\mathcal{C}}} \right)=\min \left( 1,\e^{-\beta \Delta E} \right)\,,
\eeq
where $\Delta E=E_{\mathcal{C}'}-E_{\mathcal{C}}$ is the difference between the effective classical energies of the proposed configuration $\mathcal{C}'$ and current configuration $\mathcal{C}$. 

In the absence of a quantum part to the Hamiltonian ($\Gamma=0$), not only are classical moves the only moves necessary, but they are also the only moves with a nonzero acceptance probability. In this case, the ODE algorithm automatically reduces to a classical thermal algorithm keeping the size of the imaginary-time dimension at zero ($q=0$) for the duration of the simulation. 

\subsubsection{Local swap}
A local swap is the swapping of neighboring off-diagonal operators in the sequence $S_q$.  A random pair of adjacent off-diagonal operators in the sequence is picked and swapped [as shown in Fig.~\ref{fig:updates}(b)].
If the state between $V_{i_k}$ and $V_{i_{k+1}}$ is $\ket{z}$ and is $\ket{z'}$ after the swap, then the swap involves adding an energy $E(z')$ and removing an energy $E(z)$ from the energy multiset [note that $E(z)$ and $E(z')$ may be the same].  The acceptance probability for the move is as in Eq.~(\ref{eq:Pmet}) with $M_{\mathcal{C}'}  = M_{\mathcal{C}} + \{E(z')\}-\{E(z)\}$.

\subsubsection{Block-swap} \label{sec:blockswap}
A block swap  [Fig.~\ref{fig:updates}(c)] is a local update that involves a change of the classical state $z$. Here, a random position $k$ in the sequence $S_q$ is picked such that the sequence is split into two (non-empty) parts, $S_q=S_1 S_2$, with $S_1 = V_{i_1} \cdots V_{i_{k}}$ and $S_2 = V_{i_{k+1}} \cdots V_{i_{q}}$.  The classical state $\ket{z'}$ at position $k$ in the sequence is given by
\beq
\bra{z'} =  \bra{z} S_1 =  \bra{z} V_{i_1} \dots V_{i_{k}} \ ,
\eeq
where $\ket{z}$ is the classical state of the current configuration.  The state $\ket{z'}$ has energy $E(z')$, and the state $\ket{z}$ has energy $E(z)$.  We consider a new configuration defined by
$\bra{z'} S_2 S_1 \ket{z'}$.
The multiplicity table of this configuration differs from that of the current configuration by having one fewer $E(z)$ state and one additional $E(z')$ state.  The weight of the new configuration is then proportional to $e^{-\beta M_{\mathcal{C}'}}$
where the multiset $M_{\mathcal{C}'}=M_{\mathcal{C}} + \{E(z')\} - \{E(z)\}$. 
The acceptance probability is as in Eq.~(\ref{eq:Pmet}) with the aforementioned $M_{\mathcal{C}'}$.
\subsubsection{Creation/annihilation of off-diagonal operators}
The moves presented so far have left the size of $S_q$ unchanged. 
The creation/annihilation move shown in Fig.~\ref{fig:updates}(d) has the effect of changing the value of $q$ by 2, i.e., $q \to q \pm2$, which in the transverse-field Ising model corresponds to creating or destroying off-diagonal operators $\sigma^x_j$ in pairs. We implement this via the insertion or deletion of two adjacent, identical operators.  With probability $p_{\mathrm{del}}$ (e.g., $p_{\mathrm{del}}=1/2$) we try to annihilate an adjacent pair, and with probability $1-p_{\mathrm{del}}$ we try to insert a pair.  

For pair insertion, we randomly pick an internal insertion point in the sequence (we denote this internal state by $|z'\rangle$) and a random $V$ to insert.  This adds two new energies $E(z')$ and $E(z'')$ to the multiset, where $|z''\rangle = V|z'\rangle$.  The acceptance probability for pair creation is given by  
\beq
p = \min \left(1, \frac{p_{\mathrm{del}}}{1-p_{\mathrm{del}}} \frac{N \beta^2 \Gamma^2}{(q+2)(q+3)} \e^{-\beta \Delta E}
\right)
\eeq
where as before $\Delta E=E_{\mathcal{C}'}-E_{\mathcal{C}}$ is the difference between the effective classical energies of the proposed configuration $\mathcal{C}'$ and current configuration $\mathcal{C}$ and  $M_{\mathcal{C}'} = M_{\mathcal{C}} + \{ E(z'),E(z'')\}$. 
For deletion, we randomly pick an internal point in the sequence.  If the two operators to the side of the insertion point are not identical, no deletion is performed, and the move is rejected.  If the two operators are identical, they are deleted and the relevant energies $E(z')$ and $E(z'')$ are removed from the multiplicity table. The probability of acceptance for the deletion move is
\beq
p =\min \left(1, \frac{1-p_{\mathrm{del}}}{p_{\mathrm{del}}} \frac{q(q+1)}{ \beta^2 \Gamma^2 N} \e^{-\beta \Delta E} \right) \,,
\eeq
where as before $\Delta E=E_{\mathcal{C}'}-E_{\mathcal{C}}$ and $M_{\mathcal{C}'} = M_{\mathcal{C}} - \{ E(z'), E(z'')\}$.

The size of the imaginary time dimension $q$ comes strictly from off-diagonal terms and shrinks or grows depending on the strength of the `quantum component' of the model. 
This property is expected to be heavily utilized in order to overcome the freezing of QMC algorithms in almost classical regimes.
In these regimes, $q$ is small, and the algorithm reduces to being a classical thermal algorithm\footnote{This is to be contrasted with the 
standard SSE formalism where one normally introduces an additional parameter $L$ in order to fix the size of imaginary time dimension for more efficient weight calculations. The fixing of the size of imaginary time may adversely affect the convergence of the algorithm. Here, this parameter too is spurious.}.

\subsection{Measurements}
An integral part of any QMC algorithm is the acquisition of various properties of the model such as average energy, magnetization, specific heat and correlation functions. In the ODE algorithm (as in SSE), diagonal (classical) measurements are measured differently than off-diagonal ones. 
\subsubsection{Diagonal measurements}
Diagonal operators $D$ obey $D|z\rangle=d(z)|z\rangle$ where $d(z)$ is a number that depends both on the operator and the state it acts on. Since $\langle z| D S_q |z\rangle=d(z)\langle z| S_q|z\rangle$, for any given configuration $\mathcal{C}=(|z\rangle,S_q)$, there is a contribution $d=d(z)$ to the diagonal operator thermal average $\langle D\rangle$. To improve statistics, we can also consider rotations in (the periodic) imaginary time. To do that, we may consider `virtual' block-swap moves (see Sec.~\ref{sec:blockswap}) that rotate $S_q$ and as a result also change the classical configuration from $\ket{z}$ to $\ket{z_i}$.
The contribution to the expectation value of a diagonal operator $D$ thus becomes:
\beq
d=\frac1{\mathcal{Z}} \sum_{i=0}^{q-1} d(z_i) \e^{-\beta E_{\mathcal{C}_i}} \ .
\eeq
where  $E_{\mathcal{C}_i}$ is the effective classical energy associated with configuration $\mathcal{C}_i$ whose multiset is
$M_{\mathcal{C}_i} =M_{\mathcal{C}} + \{\x(z_i)\} - \{ \x(z)\}$ (recall that $z_0 \equiv z$, so $M_{\mathcal{C}_0} = M_{\mathcal{C}}$). The normalization factor $\mathcal{Z}$ above is the sum
\beq
\mathcal{Z}=\sum_{j=0}^{q-1} \e^{-\beta E_{\mathcal{C}_j}} =\sum_j m_j \e^{-\beta E_{\mathcal{C}_j}}
\eeq
over all nonzero multiplicities $m_j$. 
In the case where $D=H_\D$ the above expression simplifies to:
\beq
d=\frac1{\mathcal{Z}}\sum_{i=0}^{q-1} \x(z_i) \e^{-\beta E_{\mathcal{C}_i}}=\frac1{\mathcal{Z}}\sum_{j} m_j \x(z_j)  \e^{-\beta E_{\mathcal{C}_j}}  \,.
\eeq
\subsubsection{Off-diagonal measurements}
%
We next consider the case of measuring the expectation value of an off-diagonal operator $V_{k}$, namely, $\langle V_{k}\rangle$.  To do this, we interpret the instantaneous configuration as follows
\bea
W_{\mathcal{C}} &=& \frac{(\beta \Gamma)^q \e^{-\beta E_{\mathcal{C}}}}{q!} \bra{z}  S_{q} \ket{z}\\\nonumber
&=&\left( \frac{\beta \Gamma} {q \e^{-\beta \Delta E}}\right) \left[\frac{(\beta \Gamma)^{q-1} \e^{-\beta E_{\mathcal{C}'}}}{(q-1)!}   \bra{z}  S_{q-1} V_{i_q} \ket{z}\right]  \,,
\eea
where $\Delta E=E_{\mathcal{C}'}-E_{\mathcal{C}}$ and $\mathcal{C}'$ is the configuration associated with the multiset $M_{\mathcal{C}'} = M_{\mathcal{C}}- \{ E(z)\}$. 
In the above form, we can reinterpret the weight $W_{\mathcal{C}}$ as contributing 
\beq
v_{k}= \delta_{k,i_q} \frac{ q \e^{-\beta \Delta E}}{\beta \Gamma} \,,
\eeq
to $\langle V_{k}\rangle$. 

As in the case of the diagonal measurements, one can take advantage of the periodicity in the imaginary time direction to improve statistics by rotating the sequence such that any of the elements of $S_q$ becomes the last element of the sequence (see Sec.~\ref{sec:blockswap}), weighted accordingly by the block-swap probability.  By doing so, $v_{k}$ becomes
\beq
v_k = \sum_i \frac{q}{\beta \Gamma} \frac{\e^{-\beta  E_{\mathcal{C}_i} } }{\sum_{j=0}^{q-1} e^{-\beta E_{\mathcal{C}_j} }} \frac{\e^{-\beta  E_{\mathcal{C}'} }}{\e^{-\beta E_{\mathcal{C}_i}  }} =
 \frac{q N_k}{\mathcal{Z}\beta \Gamma} \e^{-\beta  E_{\mathcal{C}'}} 
\eeq
where $M_{\mathcal{C}_i}  = M_{\mathcal{C}} + \{ E(z_i)\}- \{ E(z)\}$, the sum $\sum_i$ is over all rotated configurations $\mathcal{C}'$ whose $S_q$ ends with $V_{k}$, and $N_k$ is the number of times $V_k$ appears in the sequence $S_q$.
%
\subsubsection{Products of off-diagonal measurements}
%
The sampling of the expectation values of the form $\langle V_{k_1} V_{k_2}\rangle$ proceeds very similarly to the single operator case except that now both operators must appear at the end of the sequence.  The argument proceeds similarly to the single off-diagonal measurement, and we have that  the contribution to the expectation value of $\langle V_{k_1} V_{k_2}\rangle $ is
\beq
v_{k_1,k_2} = \delta_{i_{q-1},k_1} \delta_{i_{q},k_2}\frac{q(q-1)}{\beta^2 \Gamma^2} \frac{\e^{-\beta  E_{\mathcal{C}'}}}{\e^{-\beta  E_{\mathcal{C}}}} \ ,
\eeq
with $M_{\mathcal{C}'} = M_{\mathcal{C}} - \{E(z),E(z_{q-1})\}$.  As in the single off-diagonal operator case, we can use the block-swap move to alter the elements at the end of the sequence, and for each pair of adjacent operators in the sequence obtain an improved contribution 
\bea
v_{k_1,k_2} &=&\frac{q(q-1)}{\beta^2 \Gamma^2} \sum_i \frac{\e^{-\beta  E_{\mathcal{C}_i} } }{\sum_{j=0}^{q-1} e^{-\beta E_{\mathcal{C}_j} }} \frac{\e^{-\beta  E_{\mathcal{C}'_i} }}{\e^{-\beta E_{\mathcal{C}_i}  }} \nonumber\\
&=&
\frac{q(q-1)}{\mathcal{Z} \beta^2 \Gamma^2} \sum_i \e^{-\beta E_{\mathcal{C}'_i}} \,,
\eea
where $M_{\mathcal{C}_k} =  M_{\mathcal{C}} + \{ E(z_k)\}- \{ E(z)\}$, $M_{\mathcal{C}'_i} = M_{\mathcal{C}} - \{E(z),E(z'')\}$ with  $|z''\rangle = V_{k_2}|z'\rangle$ and $|z'\rangle$ is the classical state after the block swap. Similar to the single off-diagonal operator case, the sum $\sum_i$ is over all rotated configurations $\mathcal{C}'$ whose $S_q$ ends with $V_{k_1} V_{k_2}$. 

Measurements of thermal averages of products of more than two off-diagonal operators can also be derived in a straightforward manner. 
\section{Results} \label{sec:Results}
\begin{figure}[th]
\includegraphics[angle=0,scale=1,width=0.9\columnwidth]{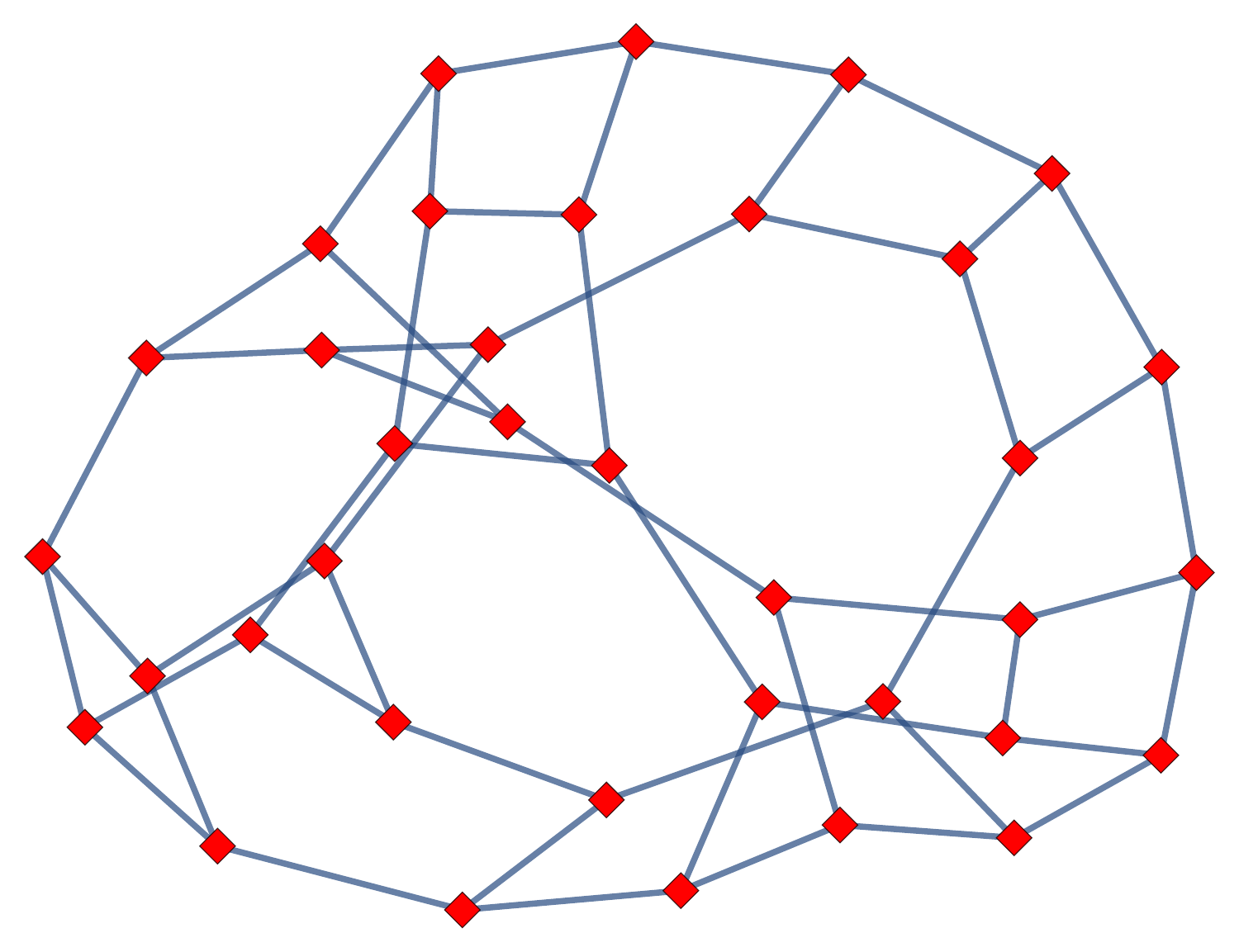}
\caption{{\bf Connectivity of a randomly generated $N=36$-spin 3-regular MAX2SAT instance.} Here the diamonds denote spins and the edges denote antiferromagnetic couplings with strength $J_{ij}=1$. Each spin is connected to three other randomly chosen spins.
}
\label{fig:instance}
\end{figure}

%
Having laid the groundwork for the ODE QMC algorithm, we present in this section some results that highlight some of its properties and advantages over existing QMC techniques, specifically a cluster-updates PIQMC algorithm\footnote{We use Wolff cluster updates~\cite{PhysRevLett.62.361} along the imaginary time direction.}. 
For benchmarking purposes, we study random 3-regular MAX2SAT instances augmented with a transverse field.   This class of instances corresponds to a particular choice of the Ising Hamiltonian given in Eq.~\eqref{eq:Hising}, whereby each spin is coupled antiferromagnetically (with strength $J_{ij} =1$) with exactly three other spins picked at random (see Fig.~\ref{fig:instance} for an illustration).  We study this class of instances as it is known to exhibit a quantum spin glass phase transition and is notoriously difficult to simulate by standard QMC techniques (see, e.g.,  Refs.~\cite{farhi:12,Liu:2015}).

\subsection{Correctness of algorithm and elastic imaginary time}

\begin{figure}[th]
\subfigure[]{\includegraphics[angle=0,scale=1,width=0.99\columnwidth]{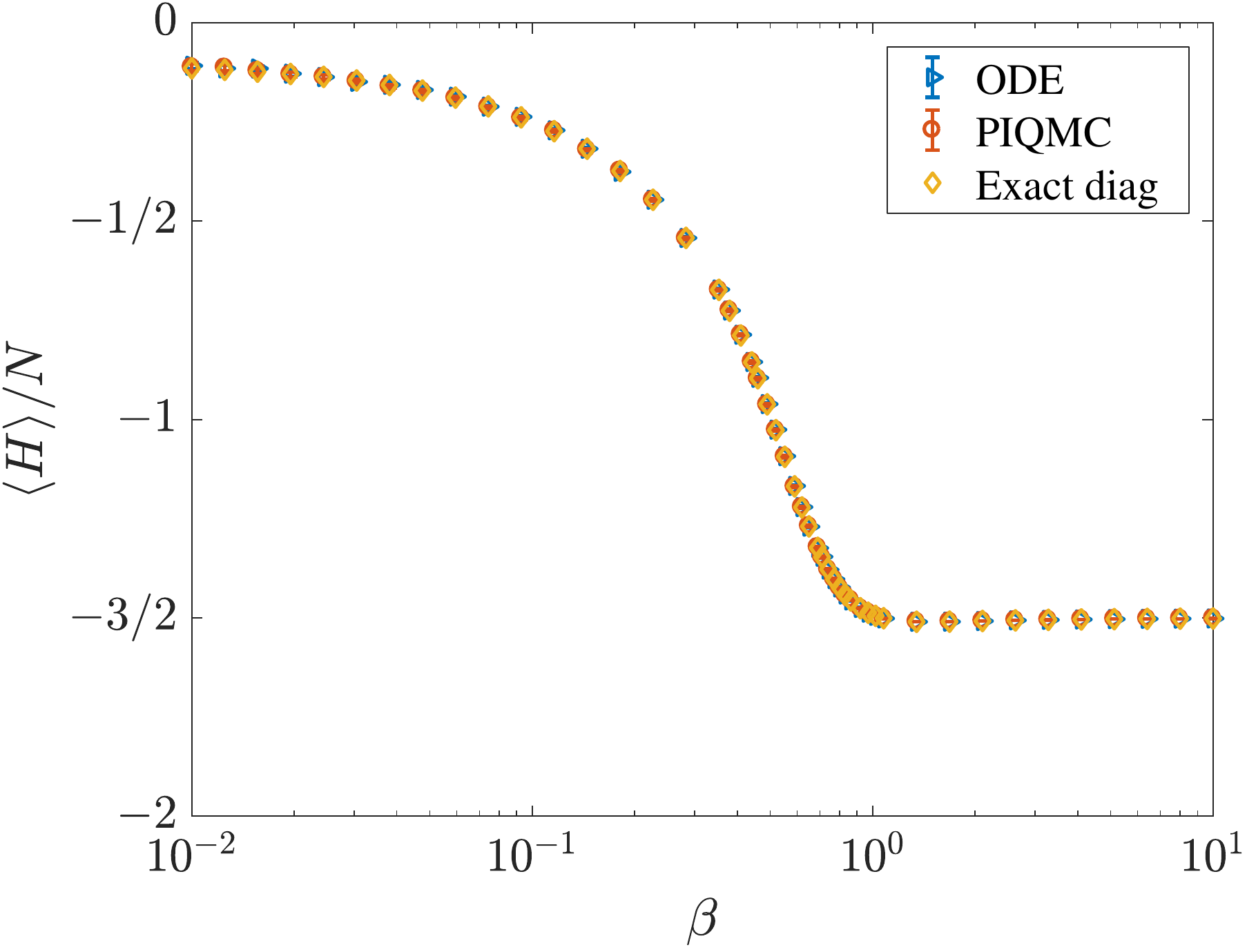}}
\subfigure[]{\includegraphics[angle=0,scale=1,width=0.99\columnwidth]{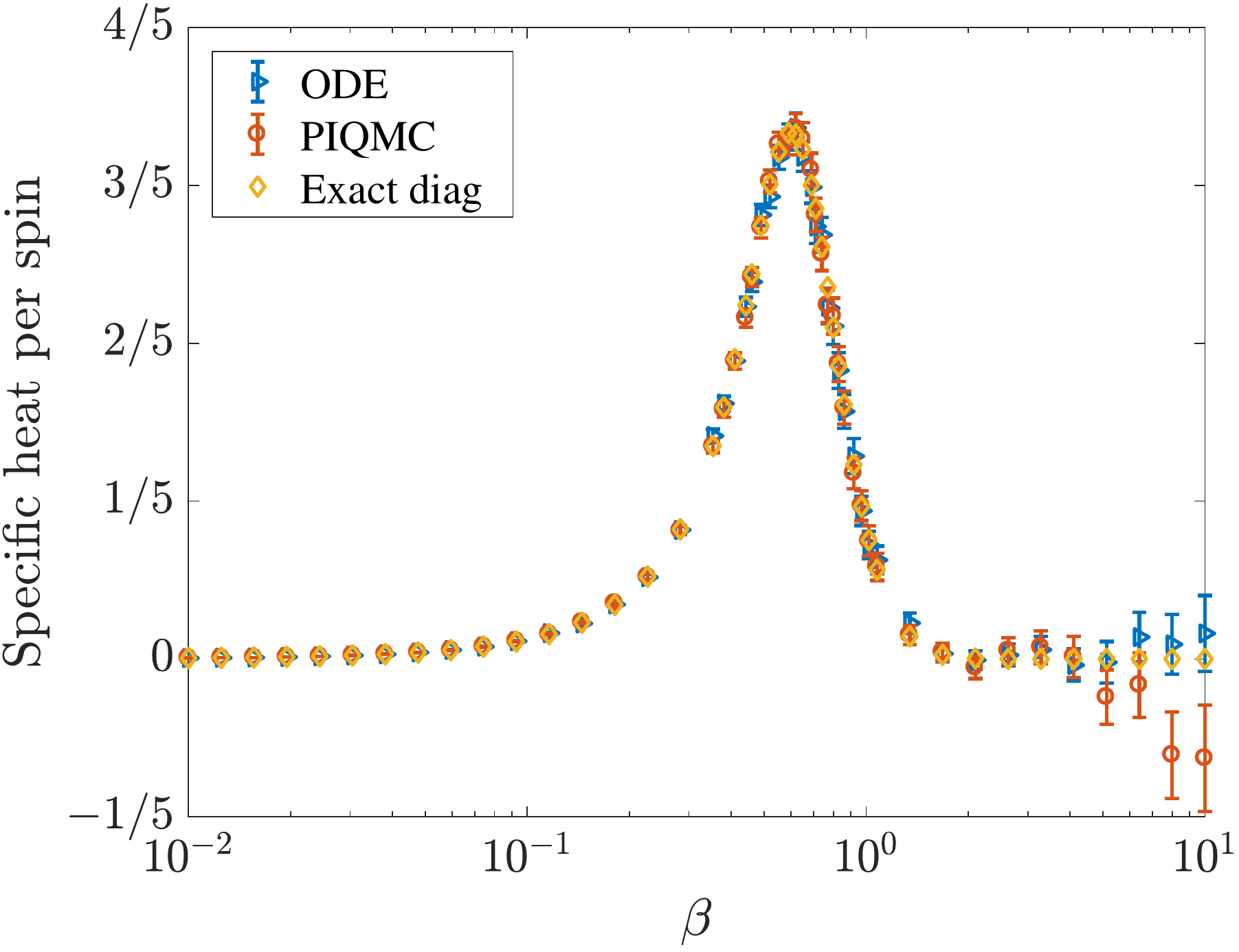}}
\caption{{\bf Agreement between ODE, PIQMC and exact diagonalization for small systems.} The thermal expectation value of the internal energy per spin $\langle H\rangle/N$ and specific heat per spin $C=\beta^2 \left(\langle H^2\rangle -\langle H\rangle^2\right)/N$ for a 3-regular MAX2SAT instance of size $N=12$ for a range of $\beta$ with $\Gamma = (10 \beta)^{-1/2}$ as calculated using ODE, PIQMC (with 5120 Trotter slices), and exact diagonalization.  Error bars correspond to $2 \sigma$ generated by performing 1000 bootstraps over the measurements.
}
\label{fig:ExactComp}
\end{figure}

As a preliminary test, we verify that we are able to reproduce the correct thermal expectation values for sufficiently small systems where comparison to exact diagonalization is feasible. An example is given in Fig.~\ref{fig:ExactComp} illustrating the excellent agreement of ODE with the exact-numerical values, even in the high-$\beta$ but low-$\Gamma$ regime where PIQMC begins to show deviations from the exact results.  Increasing the number of measurements for PIQMC rectifies this discrepancy, but the deviation already suggests that the ODE algorithm may require fewer measurements over PIQMC in the low-$\Gamma$ but large-$\beta$ regime.

\begin{figure*}[th]
\subfigure[]{\includegraphics[angle=0,scale=1,width=0.87\columnwidth]{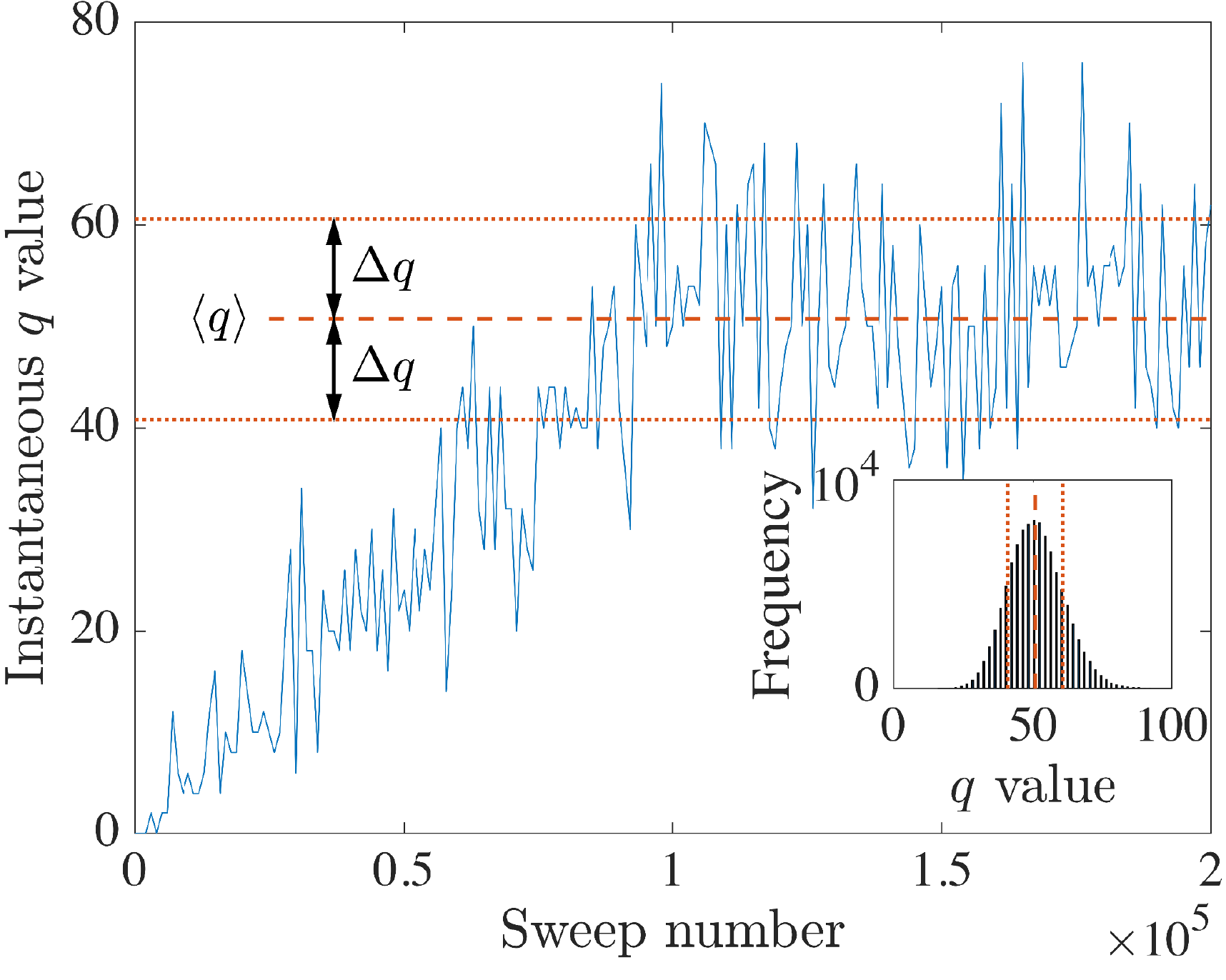}}
\subfigure[]{\includegraphics[angle=0,scale=1,width=0.9\columnwidth]{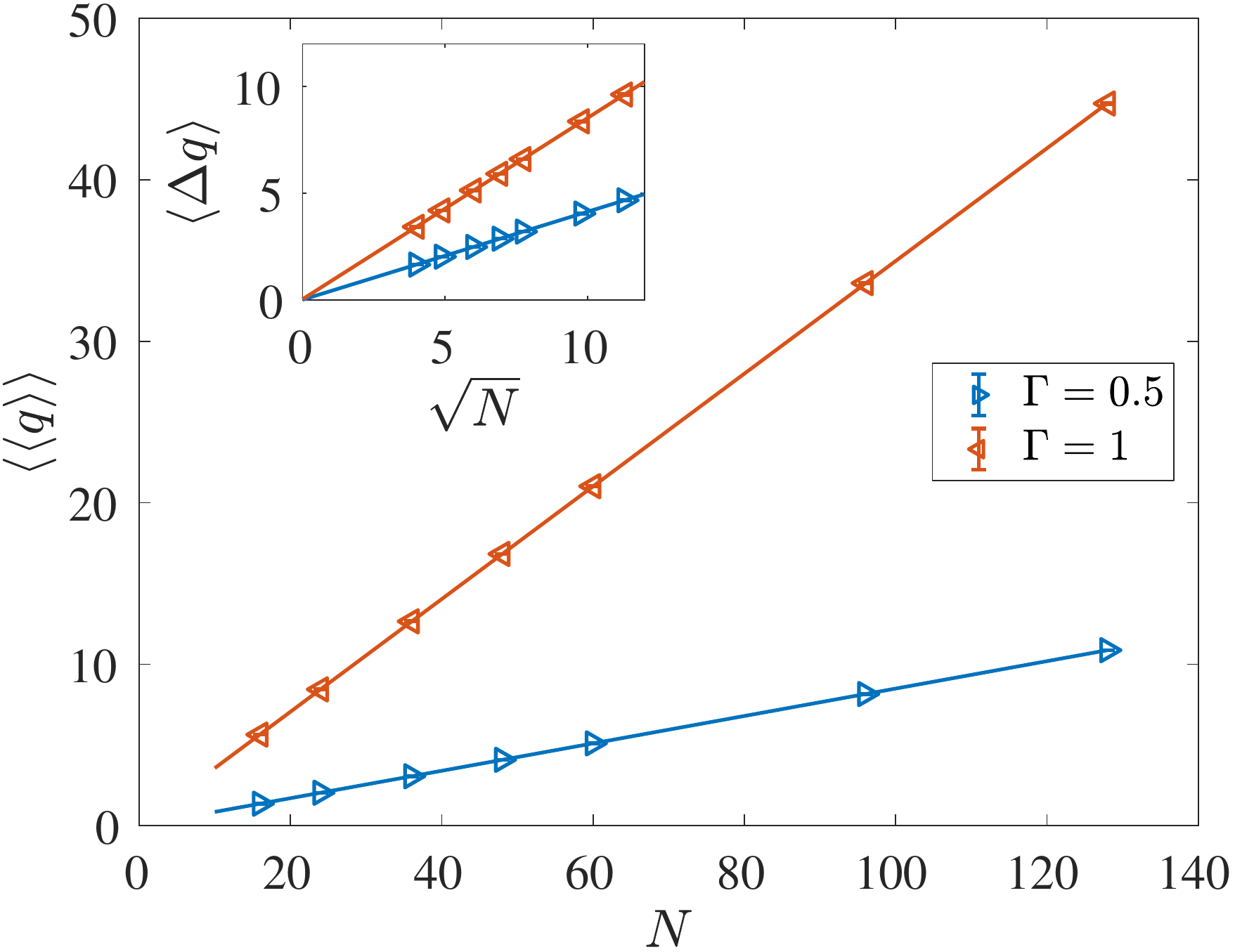}}
\subfigure[]{\includegraphics[angle=0,scale=1,width=0.9\columnwidth]{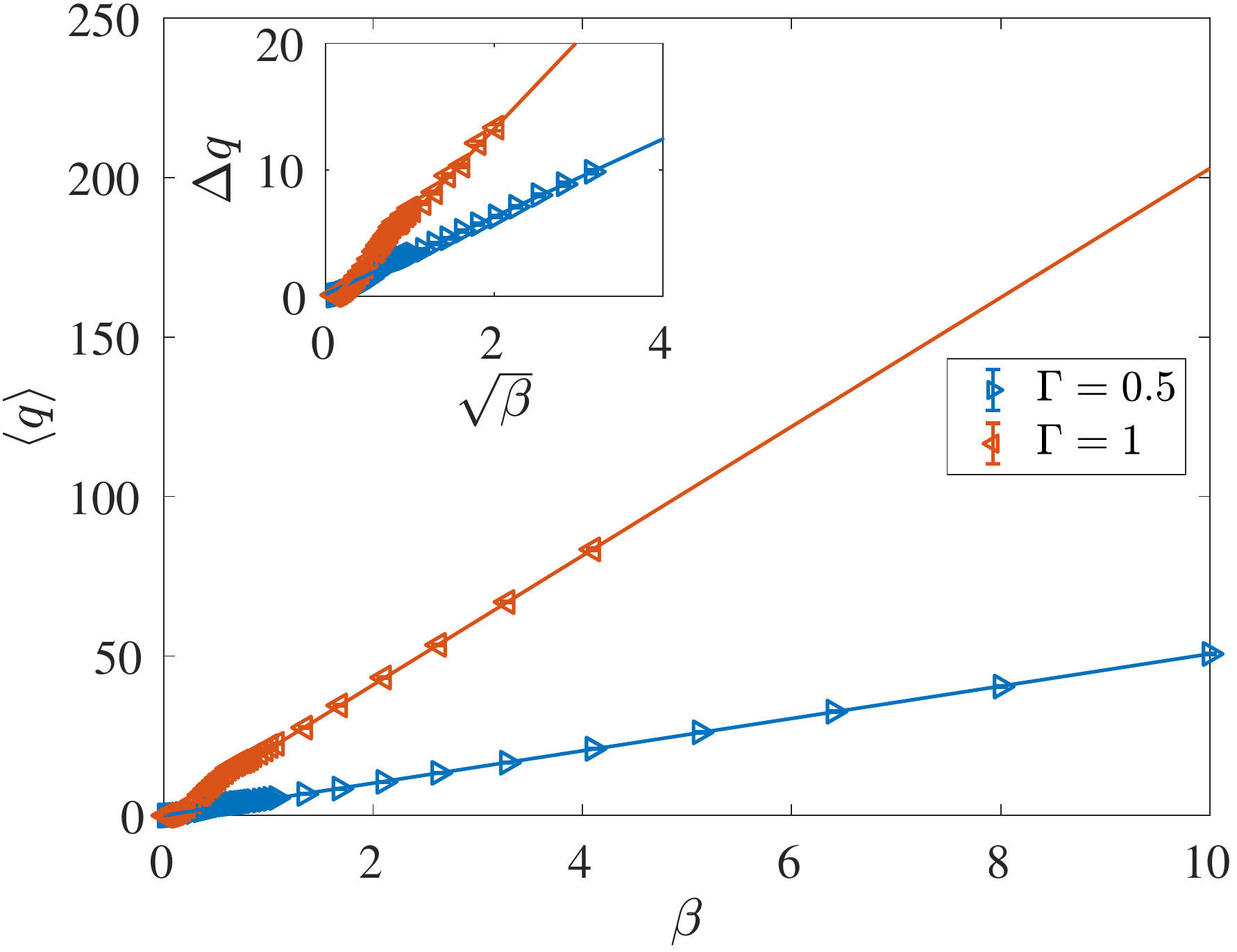}}
\subfigure[]{\includegraphics[angle=0,scale=1,width=0.9\columnwidth]{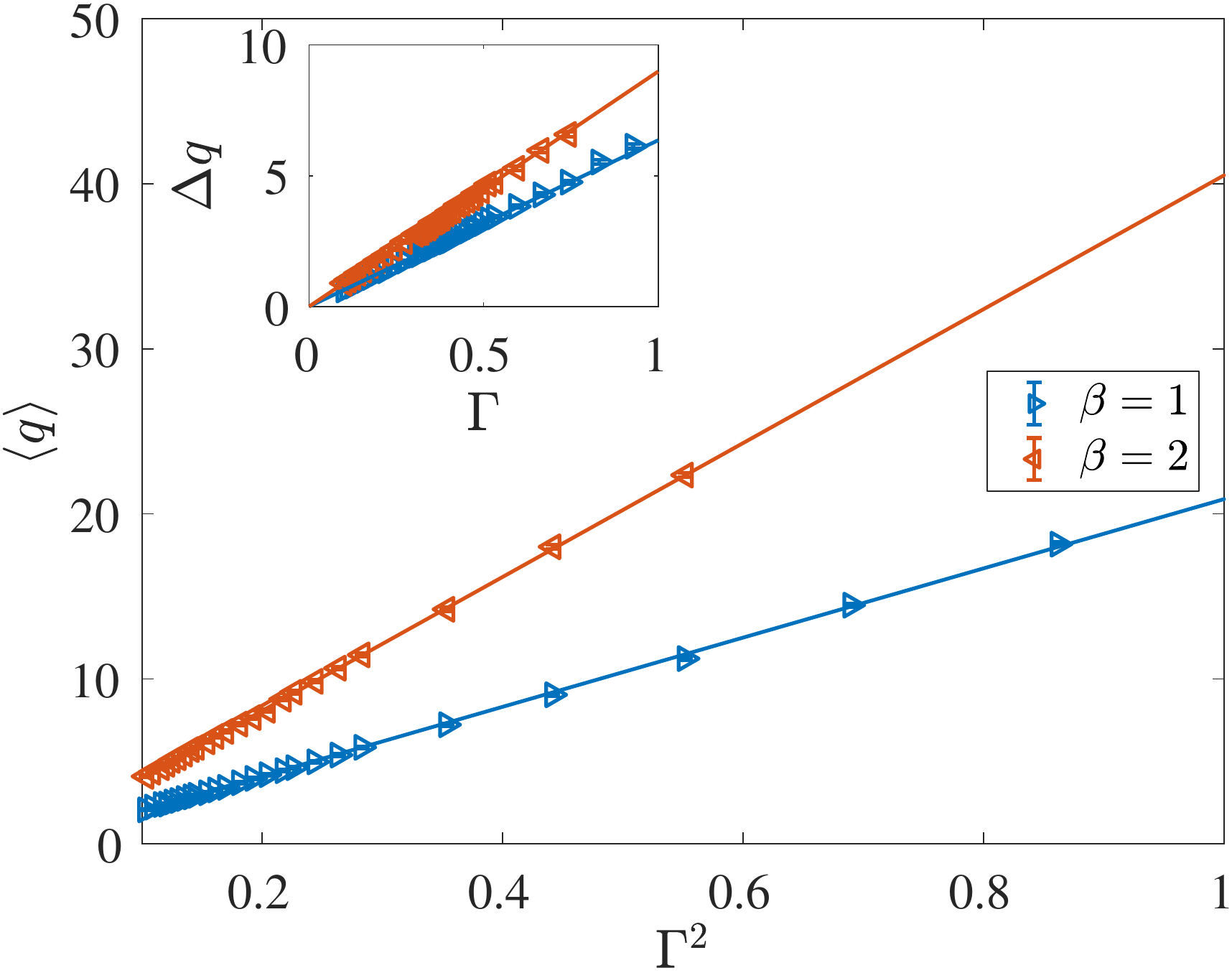}}
\caption{{\bf Size of imaginary time dimension as a function of inverse temperature, problem size and quantum strength.}  (a) Instantaneous value of $q$ as the algorithm advances, showing $q$ growing gradually from zero and then stabilizing around a mean value $\langle q\rangle$ (dashed red line) with fluctuations $\Delta q =\sqrt{\langle q^2 \rangle - \langle q \rangle^2}$ (dotted red line). (b)  $\langle q \rangle$ averaged over 48 instances (denoted $\langle \langle q \rangle \rangle$) as  a function of problem size (here, $\beta = 1$ and $\Gamma = 0.5,1$).  The inset shows $\Delta q$ averaged over 48 instances (denoted $\langle \Delta q \rangle$) as a function of $\sqrt{N}$.  (c) $\langle q \rangle$ as a function of $\beta$ for a single instance of size $N=60$ and $\Gamma = 0.5,1$.  The inset shows $\Delta q$ as a function of $\sqrt{\beta}$.  Inset shows $\Delta q$ as a function of $\sqrt{N}$.  (d) $\langle q\rangle$ as a function of $\Gamma$ for the same instance as in (b) of size $N=60$ and $\beta = 1,2$.  (b-d) The solid curves correspond to linear fits of the data points.  Error bars correspond to $2 \sigma$ generated by performing 1000 bootstraps over the measurements for (a) and (d) and over the mean from the 48 instances for (b) and (c).  
}
\label{fig:qBehavior}
\end{figure*}

We next study in Fig.~\ref{fig:qBehavior} the dependence of the average size of the imaginary time dimension, namely, $q$ on system size $N$, inverse-temperature $\beta$, and transverse field strength $\Gamma$\footnote{The warm-up of the simulations involved a linear anneal in $\beta$ from an initial value that is a factor $10^3$ smaller than the target $\beta$ to the target $\beta$.  $10^6$ sweeps are performed in total during the warm-up.  After the warm-up, $10^4$ measurements are performed, with $10^2$ sweeps between measurements to ensure the subsequent measurements are uncorrelated.}.  As was discussed earlier, the ODE QMC does not presume a-priori a size for the imaginary time dimension but rather allows it to be set dynamically during the simulation. As is shown in Fig.~\ref{fig:qBehavior}(a), as the simulation advances, the instantaneous $q$ which starts at $q=0$ gradually grows and eventually fluctuates around an average value indicating the size of the imaginary time dimension. As we expect, the average value of $q$, which we denote $\langle q \rangle$, scales linearly with $N$ and $\beta$ with fluctuations on the order of $\sqrt{N}$ and $\sqrt{\beta}$ [Figs.~\ref{fig:qBehavior}(b) and (c), respectively].
Moreover, we find that $\langle q \rangle$ does indeed grow with the quantum strength of the model. Specifically, we find it to scale quadratically with $\Gamma$ as indicated in Fig.~\ref{fig:qBehavior}(d).  

\subsection{ODE vs PIQMC}
\begin{figure*}[th]
\subfigure[]{\includegraphics[angle=0,scale=1,width=0.6\columnwidth]{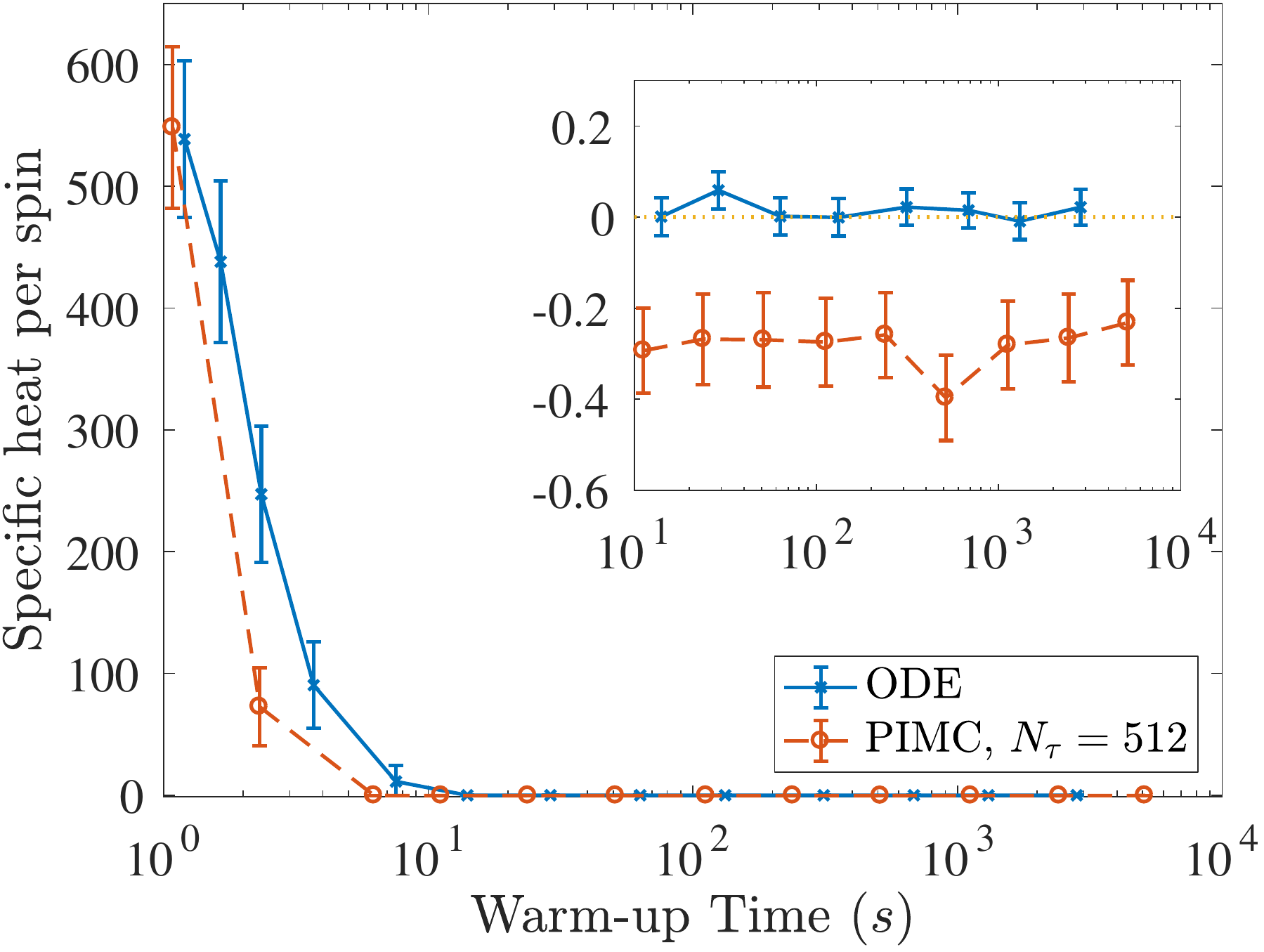}}
\subfigure[]{\includegraphics[angle=0,scale=1,width=0.6\columnwidth]{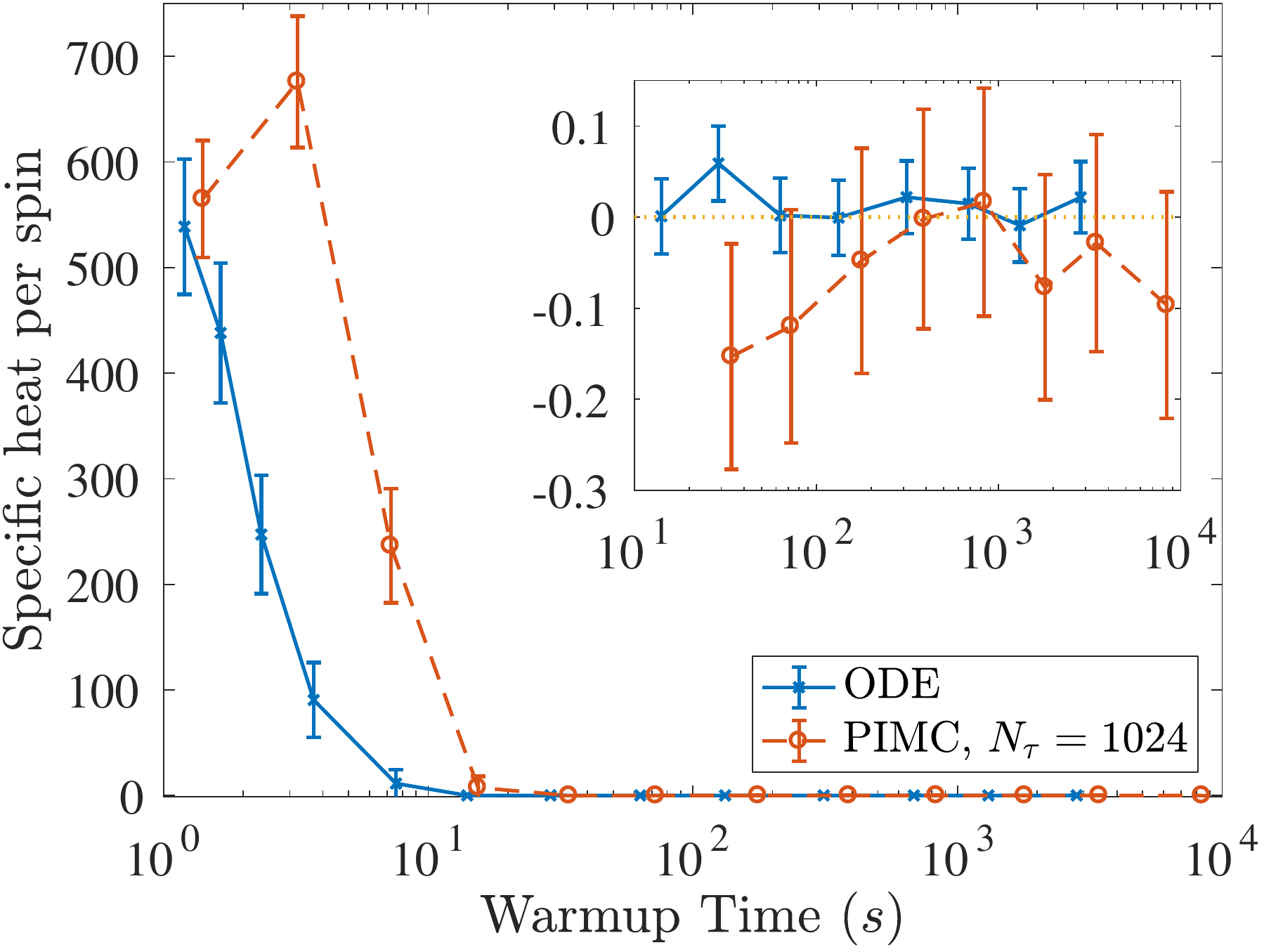}}
\subfigure[]{\includegraphics[angle=0,scale=1,width=0.6\columnwidth]{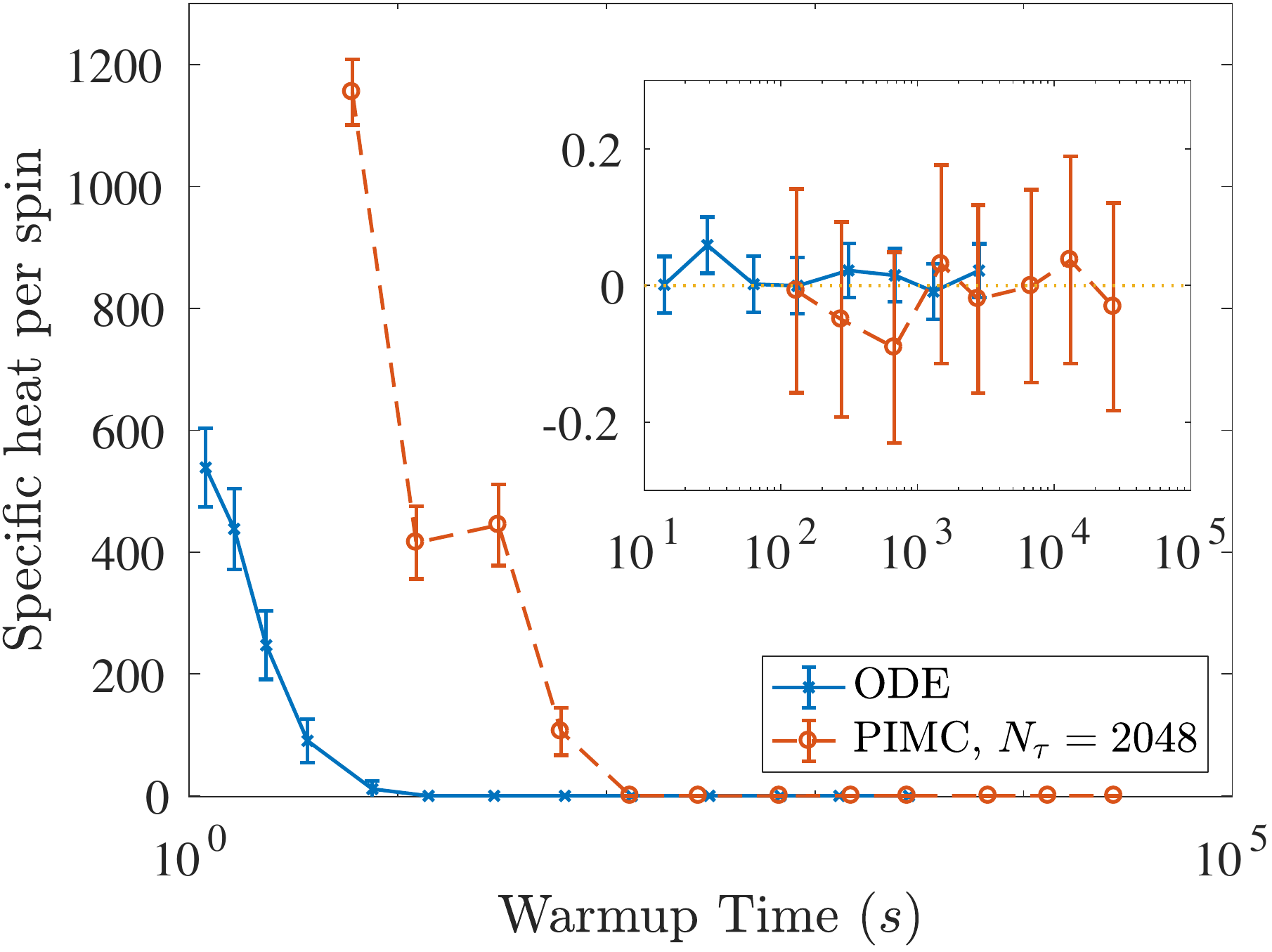}}
\caption{{\bf Performance of ODE vs PIQMC.} Required simulation time to reach a thermal state for $\Gamma = 0.1$ and $\beta = 30$.  Here, we calculate the specific heat per spin $C=\beta^2 \left(\langle H^2\rangle -\langle H\rangle^2\right)/N$.  For these values we expect the thermal state to have $\langle H^2 \rangle \approx \langle H \rangle^2$, since the thermal state should have almost all its weight on the ground state.  (a) PIQMC with 512 Trotter slices $(N_\tau)$.  (b) PIQMC with 1024 Trotter slices.  (c) PIQMC with 2048 Trotter slices.  The warm-up of both ODE and PIQMC simulations involved a linear anneal in $\beta$ from an initial value that of $0.1$ to the target $\beta$ of 30.  The number of warm-up sweeps was varied from $10^2$ to $10^6$.  After the warm-up, $100$ measurements are performed, with $10^2$ sweeps between measurements.  We ran $10^3$ independent simulations.  Error bars correspond to $2 \sigma$ generated by performing $10^3$ bootstrap over the mean values from the $10^3$ independent simulations.}
\label{fig:convergenceTime}
\end{figure*}

Since the value of $q$ determines the cost of calculating the GBWs, our results in Fig.~\ref{fig:qBehavior} indicate that the ODE algorithm can have significant advantages in the low-$\Gamma$ but large-$\beta$ regime.  For the 3-regular MAX2SAT class, this would be in the spin-glass phase, where we can expect QMC algorithms to become less efficient.  We quantify this possible advantage by comparing the performance of our algorithm against PIQMC in this regime.  In Fig.~\ref{fig:convergenceTime}, we compare the warm-up time required to reach close to the thermal state for the two algorithms.  
We observe that in order for the (discrete-time) PIQMC algorithm to achieve this, we need a sufficiently large Trotter slicing ($> 1024$), which in turn increases the time cost of performing a sweep in the simulations.  In this regard, the ODE algorithm reaches the thermal state in less computational time, with even a factor of 10 advantage when compared to PIQMC with 2048 Trotter slices.

\subsection{Quantum-classical parallel tempering\label{sec:qcpt}}
%
As we demonstrated in Sec.~\ref{eqt:Zexpansion}, the ODE~partition function decomposition naturally reduces to the classical one when the strength of the off-diagonal terms in the Hamiltonian are sent to zero.  As we show next, this 
allows us to naturally unify the classical Parallel Tempering (CPT) algorithm (also known as `exchange Monte Carlo')~\cite{hukushima:96,marinari:98b} and its quantum counterpart (QPT, see e.g., Ref~\cite{hen:11}). CPT is a refinement of the simulated annealing algorithm~\cite{kirkpatrick:83}, whereby $N_T$ replicas of an $N$-spin system at inverse-temperatures
\hbox{$\beta_1<\beta_2<\ldots< \beta_{N_T}$} undergo Metropolis spin-flip updates independently of one another and in addition, replicas with neighboring temperatures regularly attempt to swap their temperatures with probabilities that satisfy
detailed balance~\cite{sokal:97}. In this way, each replica performs a random-walk on the temperature axis, which generally allows for quicker equilibration of the system in comparison to other techniques.  Analogously in QPT, temperature is replaced by a parameter $\Gamma$ of the (quantum) Hamiltonian, e.g., the strength of the transverse magnetic field in the transverse Ising model, and each replica performs a random-walk on the $\Gamma$ axis. 

Both CPT and QPT are two widely used variations on Monte Carlo schemes but have so far been considered as separate algorithms.  
The current formulation allows to unify the two tempering algorithms in a straightforward manner. A natural generalization is to consider a tempering algorithm that traces an arbitrary curve in the this classical-quantum $\beta$-$\Gamma$ plane. This opens up the opportunity to study, e.g., certain properties of experimental quantum annealers (see for example Ref.~\cite{Dwave}) which trace such quantum-classical curves as well as to study classical-quantum optimization techniques and equilibration methods, by, e.g., looking for curves that would allow one to bypass first order phase transitions. 

If we consider replicas along a curve in the $\beta$-$\Gamma$ plane at points $\{(\beta_1,\Gamma_1),\dots, (\beta_{N_T},\Gamma_{N_T})\}$, then a parallel tempering swap probability between the $i$-th and $(i+1)$-th replica is given by:  
\bea
P &=& \min\left(1,\frac{W_{\mathcal{C}_i}(\beta_{i+1}, \Gamma_{i+1}) W_{\mathcal{C}_{i+1}}(\beta_i, \Gamma_i)}{ W_{\mathcal{C}_i}(\beta_i, \Gamma_i) W_{\mathcal{C}_{i+1}}(\beta_{i+1}, \Gamma_{i+1}) }\right) \ , 
\eea
where the above weight ratio is conveniently simplified to:
\bea
\frac{W_{\mathcal{C}_i}(\beta_{i+1}, \Gamma_{i+1}) W_{\mathcal{C}_{i+1}}(\beta_i, \Gamma_i)}{W_{\mathcal{C}_i}(\beta_i, \Gamma_i)  W_{\mathcal{C}_{i+1}}(\beta_{i+1}, \Gamma_{i+1})} &= &  \\
&& \hspace{-5cm} \left(\frac{\beta_i \Gamma_i}{\beta_{i+1}\Gamma_{i+1}}\right)^{q_{i+1}-q_i} \frac{\e^{-\beta_{i} (E'_{\mathcal{C}_{i+1}}-E_{\mathcal{C}_i})}}{\e^{-\beta_{i+1} (E_{\mathcal{C}_{i+1}}-E'_{\mathcal{C}_i})}} \nonumber \,,
\eea
where $E_{\mathcal{C}_i}$ and $E_{\mathcal{C}_{i+1}}$ are the effective classical energies of configurations $\mathcal{C}_i$ and $\mathcal{C}_{i+1}$, respectively and $E'_{\mathcal{C}_i}$ and $E'_{\mathcal{C}_{i+1}}$ are the effective classical energies of these configurations when calculated with switched $\beta$ and $\Gamma$.

\begin{figure}[htp] 
   \centering
   \includegraphics[width=0.5\textwidth]{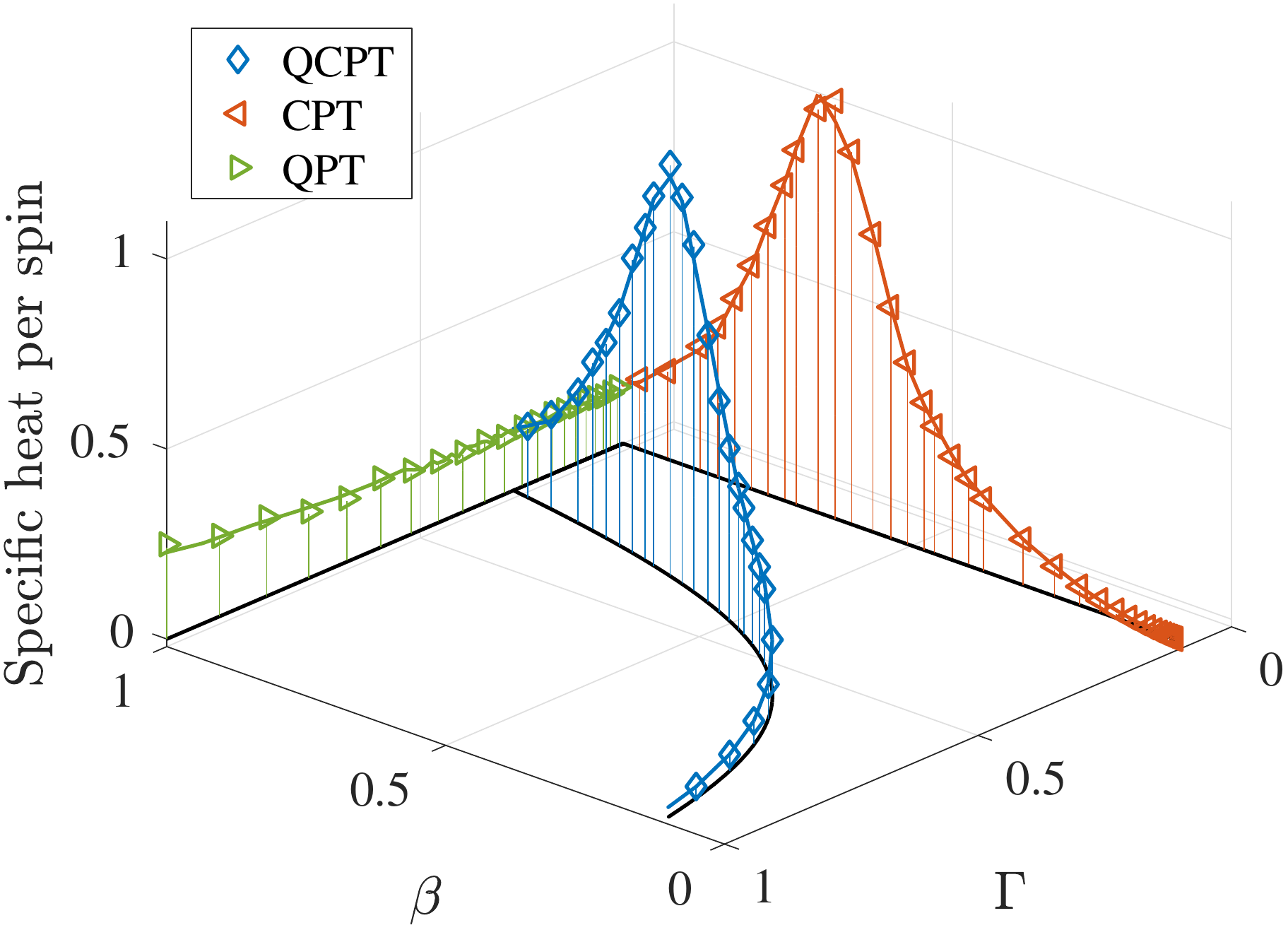} 
   \caption{{\bf Classical parallel tempering (CPT), quantum parallel tempering (QPT) and quantum-classical parallel tempering (QCPT).} Results for an instance with $N=60$ using our generalized parallel tempering algorithm along different curves in the $\beta$-$\Gamma$ plane.  Shown are the cases for CPT (red), QPT (green), and a case where $\Gamma = (10 \beta)^{-1/2}$ (blue).  To verify the accuracy of our algorithm we also show the PIQMC prediction (solid line).}
   \label{fig:PTresults}
\end{figure}

In the classical limit $\Gamma \to 0$ the ratio readily reduces to the standard CPT acceptance ratio
\beq
\frac{W_{\mathcal{C}_i}(\beta_{i+1}) W_{\mathcal{C}_{i+1}}(\beta_i)}{W_{\mathcal{C}_{i+1}}(\beta_{i+1}) W_{\mathcal{C}_i}(\beta_i) } =\e^{\Delta \beta \Delta E} \,,
\eeq
where $\Delta \beta=\beta_{i+1}-\beta_i$ and $\Delta E$ is the change in classical energy between the two configurations. Furthermore, in the case of pure quantum parallel tempering, i.e., if $\beta$ is fixed between neighboring replicas, the acceptance ratio neatly reduces to  
\beq
\frac{W_{\mathcal{C}_i}(\beta, \Gamma_{i+1}) W_{\mathcal{C}_{i+1}}(\beta, \Gamma_i)}{W_{\mathcal{C}_i}(\beta, \Gamma_i) W_{\mathcal{C}_{i+1}}(\beta, \Gamma_{i+1}) } =\left(\frac{\Gamma_{i}}{\Gamma_{i+1}}\right)^{q_{i+1}-q_i} \,.
\eeq
We show in Fig.~\ref{fig:PTresults} results for our quantum-classical parallel tempering (QCPT) algorithm along different curves in the $\beta$-$\Gamma$ plane.  The parallel tempering algorithm gives excellent agreement with numerical calculations using PIQMC with a temperature annealing protocol for every individual $(\beta, \Gamma)$ point (as in our comparisons in the previous section). The QPT algorithms used $10^5$ swap sweeps with $10$ MC sweeps per swap, whereas the PIQMC algorithm used $10^6$ sweeps and 5120 Trotter slices.  Both algorithms took $10^4$ measurements with 100 sweeps between measurements. We nicely see the precursor of the quantum phase transition in our results.

\section{Conclusions\label{sec:conclusions}}
We have developed a novel parameter-free Monte Carlo scheme designed to simulate quantum and classical many-body systems under a single unifying framework. The method is based on a decomposition of the quantum partition function that can be viewed as an expansion in the 'quantumness' of the system. 
We have argued that the classical limit of the expansion together with the elastic quantum dimension make the method suitable to simulate models that exhibit the full range of quantum and classical behavior, specifically, systems with a non-negligible classical component, which are often difficult to simulate using existing QMC techniques.  We have shown that a single weight in the proposed decomposition corresponds to infinitely many weights of the standard SSE algorithm and have demonstrated the effectiveness of our algorithm using instances from 3-regular MAX2SAT where clear advantages can be observed over PIQMC in the near-classical regime.

The feature of naturally transitioning from the quantum to the classical regime also lends itself to simulating quantum annealing \cite{Apolloni:1989fj,Apolloni:88,Somorjai:1991oq,Amara:1993rt,finnila_quantum_1994,kadowaki_quantum_1998,Brooke1999,brooke_tunable_2001,FarhiAQC:00,farhi_quantum_2001}.  Since quantum annealing processes are typically simulated by applying equilibrium QMC algorithms to a slowly changing Hamiltonian interpolating between a transverse-field initial Hamiltonian and a typically classical final Hamiltonian (see for example Refs.~\cite{troyerScience,sandvik:03,santoro,young:170503,hen:11,hen:12,farhi:12,ABLZ:12-SI,q108}; a notable exception to this approach is Ref.~\cite{sandvikAnnealingDynamics}), it has become crucially important to devise quantum Monte Carlo approaches capable of effectively simulating the full range of the quantum annealing process.  We believe our algorithm will be particularly suited for this purpose.

We also showed how the algorithm naturally unifies classical and quantum parallel tempering into a single parallel tempering process along  curves in the classical-quantum $\beta$-$\Gamma$ plane.  This highlights a key feature of our method, which is that it naturally bridges the algorithmic gap between quantum Monte Carlo and classical (thermal) Monte Carlo. This property opens up the possibility of exploring  optimal curves that speed-up equilibration in the classical-quantum plane.

We have demonstrated how the algorithm applies to the transverse-field Ising model. It would be interesting to see how it performs with respect to existing techniques on other models considered difficult to simulate. Another aspect worth studying is the existence of additional updates that are more global in nature in order to further speed up convergence. These will more likely have to be specifically tailored to the system in question. Last, methods to facilitate the evaluation of the generalized Boltzmann weights are of significance as these scale in the worst case as the square of the imaginary time dimension. More efficient methods will serve to further increase the usefulness of the ODE algorithm.  We leave the resolution of these questions for future work. 

\begin{acknowledgments}
TA was supported under ARO MURI Grant No. W911NF-11-1-0268, ARO MURI Grant No. W911NF-15-1-0582, and NSF Grant No. INSPIRE-1551064.
Computation for the work described here was supported by the University of Southern California's Center for High-Performance Computing (\url{http://hpcc.usc.edu}). 
\end{acknowledgments}

\bibliography{refs}

\appendix

\section{Notes on divided differences} \label{sec:DividedDifference}

We provide below a brief summary of the concept of divided differences which is a recursive division process. This method is typically encountered when calculating the coefficients in the interpolation polynomial in the Newton form.

The divided differences~\cite{dd:67,deboor:05} of a function $F(\cdot)$ is defined as
\beq\label{eq:divideddifference2}
F[x_0,\ldots,x_q] \equiv \sum_{j=0}^{q} \frac{F(x_j)}{\prod_{k \neq j}(x_j-x_k)}
\eeq
with respect to the list of real-valued input variables $[x_0,\ldots,x_q]$. The above expression is ill-defined if some of the inputs have repeated values, in which case one must resort to a limiting process. For instance, in the case where $x_0=x_1=\ldots=x_q=x$, the definition of divided differences reduces to: 
\beq
F[x_0,\ldots,x_q] = \frac{F^{(q)}(x)}{q!} \,,
\eeq 
where $F^{(n)}(\cdot)$ stands for the $n$-th derivative of $F(\cdot)$.
Divided differences can alternatively be defined via the recursion relations
\bea\label{eq:ddr}
&&F[x_i,\ldots,x_{i+j}] \\\nonumber
&=& \frac{F[x_{i+1},\ldots , x_{i+j}] - F[x_i,\ldots , x_{i+j-1}]}{x_{i+j}-x_i} \,,
\eea 
with $i\in\{0,\ldots,q-j\},\ j\in\{1,\ldots,q\}$ with the initial conditions
\beq\label{eq:divideddifference3}
F[x_i] = F(x_{i}), \qquad i \in \{ 0,\ldots,q \}  \quad \forall i \,.
\eeq
A function of divided differences can be defined in terms of its Taylor expansion. In the case where $F(x)=\e^{-\beta x}$, we have
\beq
\e^{-\beta [x_0,\ldots,x_q]} = \sum_{n=0}^{\infty} \frac{(-\beta)^n [x_0,\ldots,x_q]^n}{n!} \ . 
\eeq 
Moreover, it is easy to verify that
\beq \nonumber 
[x_0,\ldots,x_q]^{q+m} = \Bigg\{ 
\begin{tabular}{ l c l }
  $m<0$ & \phantom{$0$} & $0$ \\
  $m=0$ & \phantom{$0$} & $1$ \\
  $m>0$ & \phantom{$0$} & $\sum_{\sum k_j = m} \prod _{j=0}^{q} x_j^{k_j}$ \\
\end{tabular}
 \,.
\eeq
One may therefore write:
\bea
\e^{-\beta[x_0,\ldots,x_q]} &=& \sum_{n=0}^{\infty} \frac{(-\beta)^n [x_0,\ldots,x_q]^n}{n!}\\
&=&\sum_{n=q}^{\infty} \frac{(-\beta)^n [x_0,\ldots,x_q]^n}{n!} \nonumber\\
&=& 
\sum_{m=0}^{\infty} \frac{(-\beta)^{q+m} [x_0,\ldots,x_q]^{q+m}}{(q+m)!}\nonumber\\
&=&\sum_{m=0}^{\infty} \frac{(-\beta)^q}{(q+m)!} \sum_{\sum k_j = m} \prod _{j=0}^{q} (-\beta x_j)^{k_j}\nonumber
 \,.
\eea
as was asserted in the main text.

\section{Evaluation of the GBWs --- technical details} \label{sec:GBWdetails}

The basic data structures we use to store the ODE configuration $\mathcal{C}=(|z\rangle,S_q)$ are the classical configuration $|z\rangle$, which is an array of $N$ bits, and the indices for the sequence of off-diagonal operators appearing in $S_q$.
It is also useful to store
\begin{itemize}
\item
The $(q+1)$ labels/indices of the classical energies along the imaginary time dimension.
\item
The multiplicity table of classical energies $\{(m_j, E_j)\}$ counting the number of times each energy level appears 
\item 
The pyramid: an ordered set of $(q+1)(q+2)/2$ real-valued numbers. See Fig.~\ref{fig:pyr1} in the main text. 
\end{itemize}

\subsection{The pyramid}
As illustrated in Fig.~\ref{fig:pyr1} of the main text, the pyramid provides a convenient way to calculate the divided difference of 
$\e^{-\beta [E_0,\ldots,E_q]}$, or equivalently, the effective classical energy of the instantaneous configuration, namely, $\x_{(0,\ldots,q)}$ or $E_{\mathcal{C}}$.  It relies on the recursive relation given in Eq.~\eqref{eqt:EffectiveE}, namely, 
\beq \label{eqt:EffectiveE2}
\x_{(0,\ldots,q)}  = \bar{\x} -\frac1{\beta}\log \frac{2 q \sinh \beta \Delta \x}{\beta(\x_q-\x_0)} \ ,
\eeq
where 
\bea
2\bar{\x} &=& \x_{(1,\ldots,q)}+\x_{(0,\ldots,q-1)} \quad \text{and} \nonumber \\
2 \Delta \x &=&\x_{(1,\ldots,q)} - \x_{(0,\ldots,q-1)}\,, \nonumber
\eea
 with the initial conditions $E_{(i)}=E_i$. 
In the main text, we describe how the pyramid can be used to calculate the effective classical energy $E_{\mathcal{C}}$ associated with the instantaneous configuration $\mathcal{C}$.  
The base of the pyramid has $q+1$ elements, corresponding to the `initial' energies $E_{(i)}=E_{i}$ with $i=0\ldots q$. These would be the classical energies $E(z_i)$ of the intermediary classical states induced by the off-diagonal operators in $S_q$ acting on $|z\rangle$ sequentially. Let us denote this as level zero.  Level one of the pyramid, which has $q$ elements only, is now evaluated as follows.  For each element at level one, we invoke the recursion relation above using the two elements below it (see Fig.~\ref{fig:pyr1} in the main text) at level zero, i.e.,
\bea
E_{(i,i+1)}&=& \frac{E_{(i)}+E_{(i+1)}}{2} \\\nonumber
&-&\frac1{\beta}\log \frac{2\sinh \frac{\beta}{2}(E_{(i+1)}-E_{(i)})}{\beta(E_{(i+1)}-E_{(i)})}\,.
\eea
To avoid ill-defined ratios, we order the energies at level zero such that repeated values are grouped together. 
In this case, the evaluation of $E_{(i,i+1)}$ for $E_{(i)}=E_{(i+1)}$ gives $E_{(i,i+1)} = E_{(i)}$. Similarly, level two elements are calculated via
\bea
E_{(i-1,i,i+1)}&=& \frac{E_{(i-1,i)}+E_{(i,i+1)}}{2} \\\nonumber
&-&\frac1{\beta}\log \frac{4\sinh \frac{\beta}{2} (E_{(i-1,i)}-E_{(i,i+1)})}{\beta(E_{(i+1)}-E_{(i-1)})}\,.
\eea
This procedure can be continued until the top level (level $q$) of the pyramid is reached, which gives the value of $E_{\mathcal{C}}=E_{(0,\ldots.q)}$ the effective classical energy of the configuration, from which the GBW is calculated via Eq.~(\ref{eq:gbw}).
%
%

\begin{figure*}[hbp]
\includegraphics[width=0.75\textwidth]{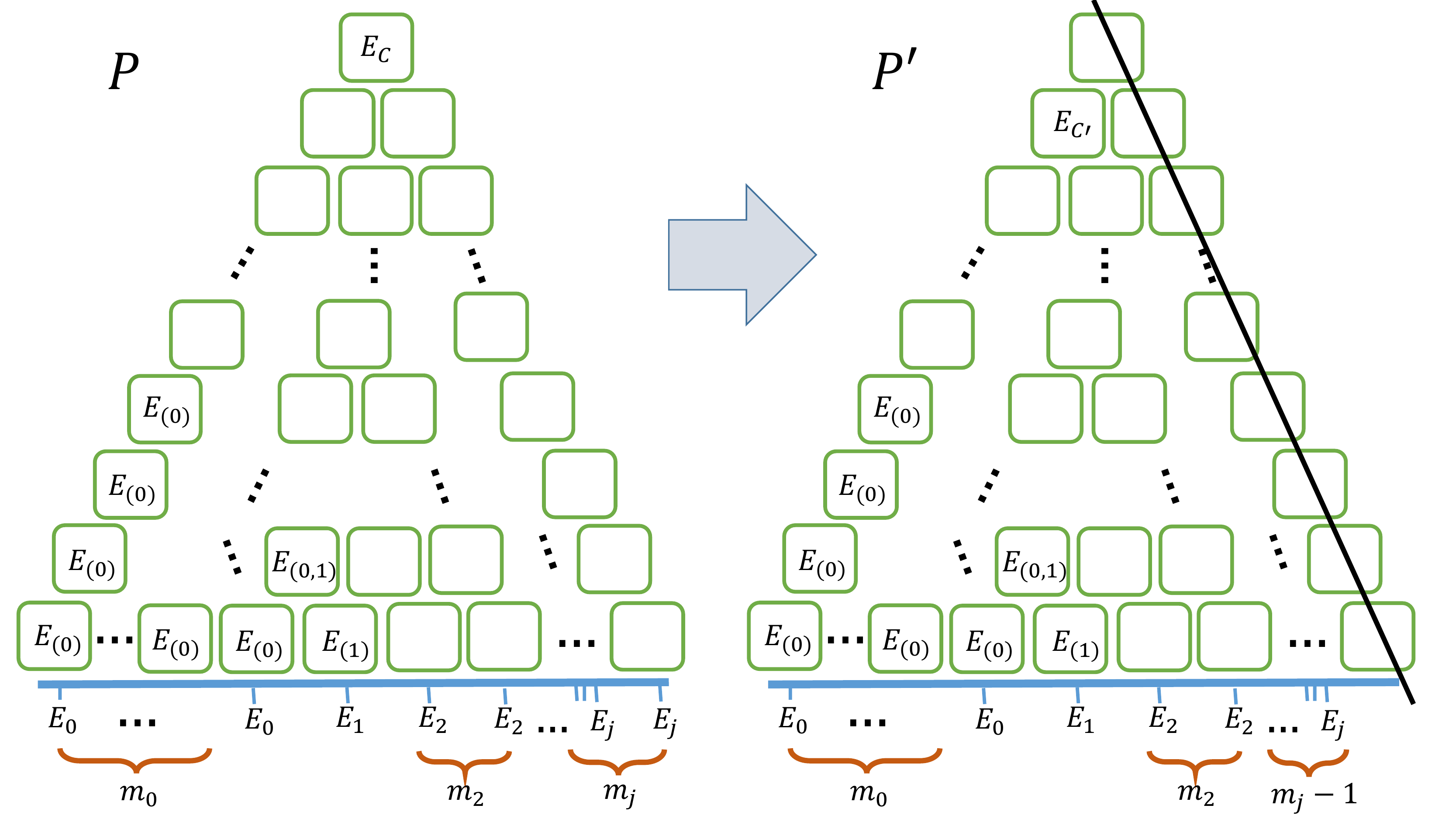}
\caption{ {\bf Removal of a single energy level}. The calculation of the effective classical energy upon a change $M_{\mathcal{C}} \to M_{\mathcal{C}} -\{E(z')\}$ requires no calculations if the removed energy level is from the edge of the pyramid.  In order to illustrate this, let us assume that the energy $E(z') = E_j$ (with multiplicity $m_j$) occurs at the right-most edge of the pyramid $P$, as depicted in the left panel.  The removal of this energy from the base of the pyramid $P$ eliminates the right-most element for all higher levels of the pyramid, including the top-most element, as depicted in the right panel.  The new top element of the pyramid $P'$ is the element labelled $E_{\mathcal{C}'}$, which requires no further calculation since it is inherited from the pyramid $P$.} 
\label{fig:removal}
\end{figure*}

\subsection{Virtual vs actual moves}
Naively, calculating the value of a GBW, or equivalently the effective classical energy of a configuration, requires $(q+1)(q+2)/2$ operations as the number of blocks in the pyramid. However, small changes to an already evaluated pyramid generate a new pyramid whose GBW is easier to evaluate. For instance, the GBW associated with the removal of a single energy value $M_{\mathcal{C}} \to M_{\mathcal{C}}-\{E(z')\}$ requires no calculations if the to-be-removed energy $E(z')$ appears at an outer edge of the base level. This is illustrated in Fig.~\ref{fig:removal}. (It can be shown that similar tricks may be applied even if the energy level to be removed is from the `bulk' of the pyramid.) Similarly, the addition of a single energy level often requires only $O(q)$ operations.

\begin{figure*}[hbp]
\includegraphics[width=0.75\textwidth]{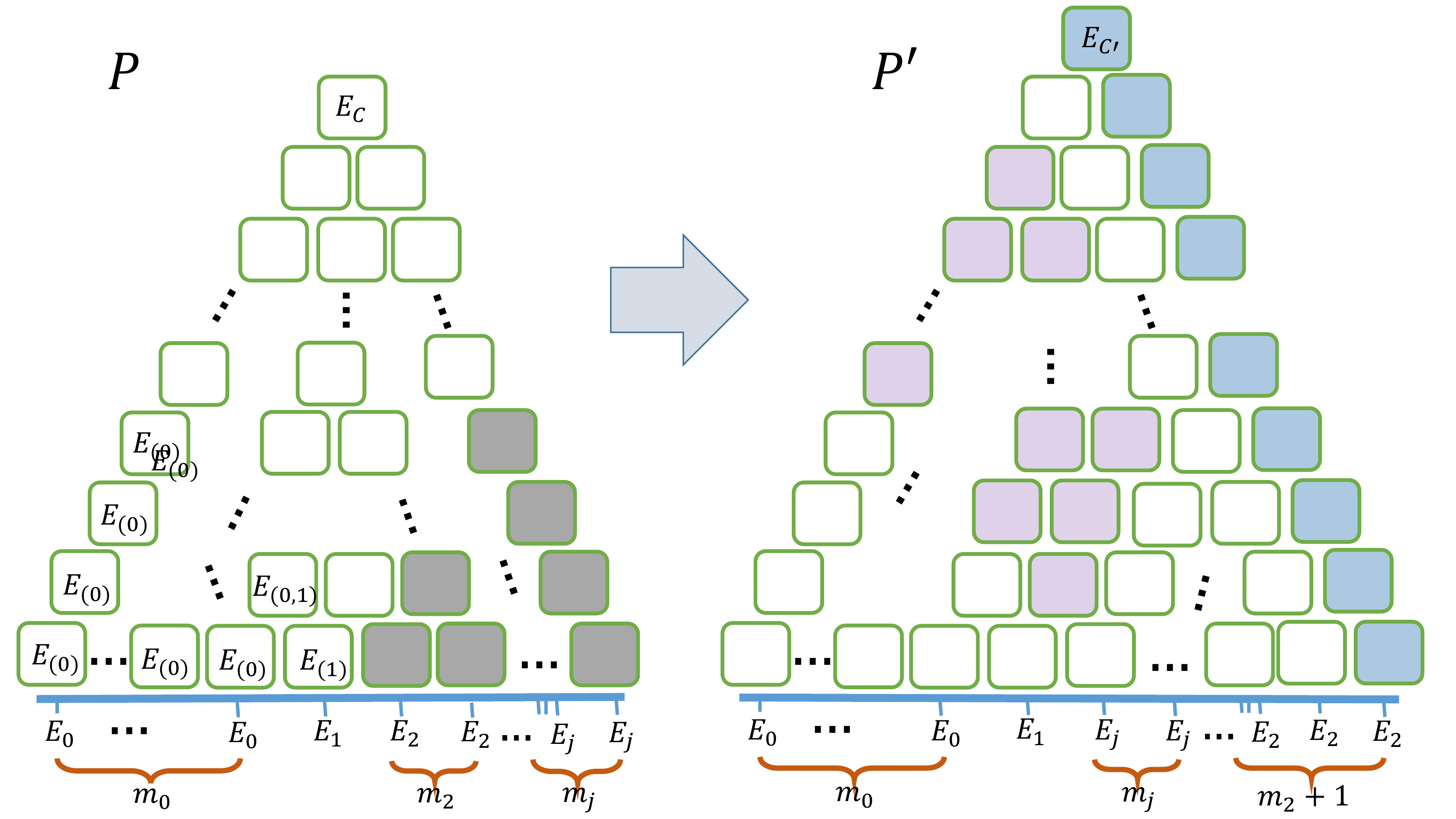}
\caption{ {\bf Manipulations of the pyramid for a virtual move}.  The multiplicity table is such that the energy $E_i$ appears $m_i$ times (the multiplicity of the energy $E_i$), with $\sum_i {m_i} = q+1$.  Let us consider that a proposed move changes the multiplicity of the energy $E_2$, i.e., $m_2 \to m_2 +1$.  In order to determine the pyramid $P'$ associated with this change, we perform the following manipulations to the original pyramid $P$.  The sub-pyramid associated with $E_2, \dots E_j$ can be `flipped' such that the energy $E_2$ appears at the edge of $P$ (dark-colored blocks on left panel).  Because this manipulation does not change the base of the sub-pyramid nor that of the entire $P$, the values at the top level of sub-pyrmaid and $P$ remain unaffected (the unaffected blocks are the empty blocks on the right panel). However, this move does change the values of other elements in $P$ (purple colored blocks), but we will not need to calculate them.  Because $E_2$ now appears at the edge of $P$, introducing an additional $E_2$ to the base of $P$ (to generate $P'$) requires us to recalculate the new elements that appear at the edge of $P'$, namely only $O(q)$ operations.} 
\label{fig:swap}
\end{figure*}

A key property of the divided difference of a function is that it is invariant under reordering of the input values.  In the context of our `pyramid scheme' of calculating the divided difference, this means that while the ordering of groups of identical energies in the multiplicity table changes the intermediate values of the pyramid, it does not change the value of the top-most level of the pyramid.  Therefore, by manipulating the ordering of the energies such that local changes to the multiset of energies (as occurs for the local swap, block swap and annihilation/creation moves described in the main text) occur at the edges of the pyramid, it is possible to minimize the number of computations needed to determine the top-most level of the pyramid from $O(q^2)$ to $O(q)$ or $O(1)$.  This allows us to calculate the weights of proposed changes more efficiently than recalculating the entire pyramid, although this procedure may leave some elements of the pyramid undefined.  These virtual moves are highly useful both for updates as well as in measurement steps where virtual rotations of $S_q$ are useful. Only if the move is accepted do we need to calculate these `missing', or unevaluated, elements of the pyramid.  We call this process a `virtual move.'  We illustrate one such procedure in Fig.~\ref{fig:swap}.  

The computational complexity associated with calculating the changes to the effective classical energy (equivalently, the change to the GBW) due to the local updates and measurements discussed in the main text is summarized in Table~\ref{table:complex}.

\begin{table*}[htp]
\centering
\begin{tabular}{|l||l|c|}
\hline
Update  & Change to the energy & Computational \\
  & multiset $M_{\mathcal{C}}$  & complexity (worst case) \\
\hline \hline
Local swap & $M_{\mathcal{C}} \to M_{\mathcal{C}} +\{ E(z_i) \} -\{ E(z_j) \}$ & $O(1)$ \\\hline
Block swap & $M_{\mathcal{C}} \to M_{\mathcal{C}} +\{ E(z_i) \} -\{ E(z_j) \}$ & $O(1)$  \\\hline
Pair creation & $M_{\mathcal{C}} \to M_{\mathcal{C}} +\{ E(z_i),E(z_j) \}$ & $O(q)$  \\\hline
Pair annihilation & $M_{\mathcal{C}} \to M_{\mathcal{C}} -\{ E(z_i),E(z_j) \}$ & $O(q)$  \\\hline
$\langle V_k \rangle$ measurement& $M_{\mathcal{C}} \to M_{\mathcal{C}} -\{ E(z) \}$ & $O(q)$  \\\hline
\end{tabular}
\caption{{\bf Computational complexity of virtual updates for local changes to the multiplicity table.} While naively the calculation of a GBW required $O(q^2)$ operations, changes to the GBW can carried out much more efficiently if the multiplicity table is only locally perturbed.}
\label{table:complex}
\end{table*}

\subsection{Precision issues}
The calculation of the effective classical energy $E_{\mathcal{C}}$ and GBWs using the recursion scheme described above may require recursive operations on pairs of numbers of approximately equal magnitude whose difference is an order of magnitude closer to zero.  If we restrict ourselves to a fixed bit-precision representation of the numbers, the calculated difference may be erroneous because of truncation errors.  In order to avoid this problem, it is  necessary to check periodically whether an increase in the bit-precision used changes the result.  In our simulations, we initially use 53 bits for the significand (also known as mantissa), which is the number used for `double' precision in the ANSI/IEEE-754 standard, but increase the number of bits by factors of 1.2 when necessary using the GNU Multiple-precision Binary Floating-point Library with Correct Rounding library~\cite{MPFR}.

\section{\label{sec:npq} Calculation of $N_p(q)$ [Eq.(\ref{eq:npq})]}
The calculation of $N_p(q)$, the number of $S_q$ sequences comprised of $q$ off-diagonal $\sigma^x_i$ operators ($i=1\ldots N$) such that each operator appears an even number of times is carried out as follows~\cite{npqRef}. By definition, 
\beq
N_p(q) = \sum_{\sum_{i=1}^ N k_i=q, k_i \text{even}} {n \choose k_1 k_2 \ldots k_N} \,,
\eeq
where $k_i$ is the number of times that operator $\sigma^x_i$ appears in the sequence. We note that ${n \choose k_1 k_2 \ldots k_N}$ is the is the coefficient of $x_1^{k_1} x_2^{k_2} \cdots x_N^{k_N}$ in the expansion of $\left(x_1+x_2+\ldots+x_N \right)^q$.
The sum of all these coefficients is obtained by substituting $x_1=x_2=\ldots=x_N$.
To eliminate odd powers $k_1$, we can consider the expansion of 
\beq
\frac1{2} \left[ \left( x_1+x_2+\ldots +x_N \right)^q+\left( -x_1+x_2+\ldots +x_N \right)^q \right] \,.
\eeq
Continuing this way, we eventually arrive at
\beq
N_p(q) = \frac1{2^N} \sum_{t_i=0,1} \left( (-1)^{t_1} +(-1)^{t_2} + \ldots +(-1)^{t_N}\right)^q \,,
\eeq
which can be further simplified to
\beq
N_p(q)  = \frac1{2^N}\sum_{k=0} {N \choose k} (N-2k)^q\,.
\eeq

\end{document}